\documentclass[11pt,english]{article}
\usepackage[utf8]{inputenc}
\usepackage{amsmath}
\usepackage{amssymb}
\usepackage{cite}
\usepackage{ragged2e,amsfonts,graphicx,caption,subcaption,xcolor}
\usepackage[left=2cm,right=2cm,top=3cm,bottom=3cm]{geometry}
\usepackage{hyperref}
\hypersetup{colorlinks=false, linkbordercolor={white}}
\makeatletter

\newcommand{\sech}{\mbox{sech}}
\usepackage{babel}
 
\begin{document}
\begin{titlepage}
\thispagestyle{empty}

\bigskip

\begin{center}
\noindent{\Large \textbf {Consistency Conditions for $p$-Form Fields Localization on Codimension two Braneworlds}}\\

\vspace{0,5cm}

\noindent{L. F. F. Freitas${}^{a}$\footnote{e-mail: luizfreitas@fisica.ufc.br}, G. Alencar${}^{a}$\footnote{e-mail: geova@fisica.ufc.br} and R. R. Landim${}^{a}$\footnote{e-mail: renan@fisica.ufc.br}}

\vspace{0,5cm}
 
{\it ${}^a$Departamento de Física, Universidade Federal do Ceará-
Caixa Postal 6030, Campus do Pici, 60455-760, Fortaleza, Ceará, Brazil. 
 }

\end{center}

\vspace{0.3cm}

\begin{abstract}\noindent
Recently, in (Eur.Phys.J.C 80 (2020) 5, 432), the present authors obtained general stringent conditions on the localization of fields in braneworlds by imposing that its zero-mode must satisfy Einstein's equations (EE). Here, we continue this study by considering free $p$-form. These fields present an on-shell equivalency relation between a $p$-form and a $(D-p-2)$-form, provided by Hodge duality (HD). This symmetry will impose a new consistency condition, namely, confinement of a $p$-form must imply the localization of its dual. We apply the above conditions to $6$D braneworlds. With this, we find that in global string-like defects, for example, the $1$-form has a normalizable zero-mode. By using the HD, we show that its bulk dual $3$-form also has a normalizable zero-mode, making the confinement consistent with HD. However, these solutions cannot be made consistent with EE, therefore, these fields must be ruled out. In fact, by imposing both conditions, only the scalar and its dual field can be consistently localized. In this way, all the literature so far in which the free $1$-form is localized in codimension two models should be reviewed. These results also point to the fact that the symmetries of the fields can be used to verify the consistency of their localization and even prohibit it. 
\end{abstract}
\end{titlepage}

\tableofcontents

\section{Introduction}\label{Sec-1} 

In braneworld context, our $4$-dimensional spacetime is regarded as a hypersurface ($3$-brane) embedded in a higher dimensional bulk. Among the most popular models are those proposed by Randall-Sundrum (RS) \cite{RS1, RS2}. These models became very attractive because gravity can be confined on a delta-like $3$-brane, and thus, Newton's law of gravitation can be recovered. In addition to gravitational aspects, other important points related to the Standard Model (SM) fields can also be studied. Although RS considered all the SM fields previously confined on the $3$-brane, further studies have shown that most of these fields, propagating freely on the bulk, are not localized on the brane \cite{Chang, Barut}. This fact gave rise to another line of study related to the localization of the Standard Model fields in braneworld scenarios.

After the success of the RS models, other proposals of braneworlds with localized gravity have been presented. In $5$D for example, smooth versions (thick branes) of RS-II  were proposed, where the $3$-brane is generated by a scalar field propagating on the bulk \cite{Gremm1, Kehagias}; thick brane models with inner structure \cite{Bazeia0, Bazeia1}; branes generated by purely geometric quantities \cite{Cendejas}; or still, braneworlds in a cosmological context, where the $3$-brane has a \textit{Robertson-Walker} metric \cite{Liu0, Guo}. In addition to these, other solutions in higher-dimensional scenarios were proposed. The $6$D models, for example, deserve special attention. The reason is that this case also has analytical solutions for the metric generated by topological defects \cite{Gherghetta02}. The most common in the literature are those where the $3$-brane are generated by string-like or vortex defects. Generally, the metric for such cases is given by
\begin{equation}\label{EQ-1-01}
ds^{2}=g_{MN}dX^{M}dX^{N}=\alpha^{2}(r)\hat{g}_{\mu\nu}(x)dx^{\mu}dx^{\nu}+\gamma^{2}(r)d\theta^{2}+dr^{2},
\end{equation}
where $r\in[0,\infty)$ and $\theta\in[0,2\pi)$ are the extra dimensions. The above metric is considered the {\it vacuum solution} for the Einstein's equations obtained from the action
\begin{equation}\label{EQ-1-02}
S_{\mbox{grav.}}=\int d^{4}xdrd\theta\sqrt{-g}\left[\frac{1}{2\kappa^{2}}\left(R-2\Lambda\right)+\mathcal{L}_{b}\right],
\end{equation}
where $g$ is the determinant of $g_{MN}$ and $\mathcal{L}_{b}$ is the matter source of the brane. Reference \cite{Cohen} was the first to present a metric with the above features by assuming $\Lambda=0$. The authors showed that, although their solution had a naked singularity at $r=0$, a consistent gravitational theory on the brane could be obtained. Soon after, in Refs. \cite{Gregory,Gherghetta}, the authors found a non-singular solution valid for the exterior of a string-like topological defect. For this, they included a negative cosmological constant in the bulk ($AdS_{6}$). The metric for this model is given by (\ref{EQ-1-01}) with
\begin{equation}\label{EQ-1-03}
\alpha^{2}(r)=\exp\left(-kr\right),\ \ \ \ \ \ \gamma^{2}(r)=R^{2}_{0}\alpha^{2}(r).
\end{equation}
Other solution was obtained in Refs. \cite{Carlos, Silva}. In this model, the metric is valid outside and inside the string-like defect. The warp factors $\alpha(r)$ and $\gamma(r)$ in (\ref{EQ-1-01}) are given by,
\begin{equation}\label{EQ-1-04}
\alpha^{2}(r)=\exp\left[-kr+\tanh\left(kr\right)\right],\ \ \ \ \ \ k^{2}\gamma^{2}(r)=\tanh^{2}\!\left(kr\right)\alpha^{2}(r).
\end{equation}
Unlike the previous cases, this metric provides a natural thickness for the string-like defect ($3$-brane). Beyond this, the bulk geometry is an asymptotically $AdS_{6}$ space, which is a desirable characteristic in the study of gravity and matter fields localization. In addition to these string-like models, other, also in $6$D, were proposed. In Refs. \cite{Kaloper2004, Flachi}, the $3$-brane is generated by the intersection of two delta-like $4$-brane. The metric for this model is given by,
\begin{equation}\label{EQ-1-06}
ds^{2}=\frac{1}{\left(1+k_{1}|y|+k_{2}|z|\right)^{2}}\left[\eta_{\mu\nu}dx^{\mu}dx^{\nu}+dy^{2}+dz^{2}\right],
\end{equation}
Here, we have also an asymptotically $AdS_{6}$ bulk space, with the two extra dimensions, $y$ and $z$, infinitely large. There are yet other models in higher-dimensional configurations. For example, braneworlds generated by the intersection of an arbitrary number of delta-like branes \cite{Arkani02}, beyond other proposals \cite{Choudhury, Kim, Corradini, LiuFu}.

For all the models mentioned above, confinement of other fields, beside the gravitational one, is always an important point to be verified \cite{Daemi, Koley, Ringeval, Landim, Melfo}. When we talk about confinement, all studies are based on the {\it finite integral} argument and this approach is used for any field. This method relies on the possibility of factorizing the action
\begin{equation}
S=\int d^{4}x d^{D-4}z\sqrt{-g^{(D)}}\mathcal{L}_{\mbox{(matter)}}^{(D)},\label{EQ-1-07}
\end{equation}
into an effective action on the $3$-brane and an integral in the coordinates of the extra dimensions, {\it i.e.},
\begin{equation}
S=\int d^{D-4}z f(z)\int d^{4}x \sqrt{-g^{(4)}}\mathcal{L}_{\mbox{(matter)}}^{(4)}=K\int d^{4}x \sqrt{-g^{(4)}}\mathcal{L}_{\mbox{(matter)}}^{(4)}.\label{EQ-1-08}
\end{equation}
Thus, the theory will be well-defined, {\it i.e.}, the field will be confined on the brane, when the integral $K$ in the extra coordinates is finite. This argument is commonly used as a sufficient condition to affirm that a field is localized. A particular class of fields that we can highlight are the $p$-form fields. Among them, the $U(1)$ vector field ($1$-form) and the Kalb-Ramond field (two-form) has an important status. The study about the confinement of these two fields and other $p$-form was already widely performed in the literature\cite{Ghoroku, Alencar02, Chumbes, Alencar03, Cruz01, Cruz02, Freitas, Costa01, Costa02, Zhao02, Giovannini, Alencar2010a, Alencar2010b, Jardim02, Fu01, Fu02, Fu03, Fu04, Chen, Sousa}. As well as other related issues, such as: $p$-form fields has been used to provide the stabilization of the radius in RS-I model \cite{Rama}; or, to introduce torsion in RS-I scenarios \cite{Mukhopadhyaya01}; or still, to generate inflation or gravitational waves \cite{Germani, Kobayashi}; among other issues \cite{G.Landim01, G.Landim03, G.Landim04}. In this context, Refs. \cite{Duff, Hahn2001} performed a very interesting discussion for condimension one RS scenarios. They showed that the free $p$-form field confinement must satisfy not only the finite integral requirement, but also a symmetry provided by the Hodge duality transformation. In summary, they claimed that if a specific $p$-form field is confined in a particular braneworld, its dual $(D-p-2)$-form field also must be confined. In this way, it is possible to obtain other confined $p$-form fields only by symmetry requirement. Ref. \cite{Duff} goes further and, by using Einstein's equations, the authors obtain an additional requirement for the localization in codimension one RS models to be consistent. In a recent study performed by the present authors in Ref. \cite{Freitas02}, we explored the Einstein's equations to get some general conditions that any Standard Model field must satisfy to provide a consistent localization on the brane.

In this direction, we propose to study the localization of a free arbitrary $p$-form field in codimension two braneworld models. We will show that the Hodge duality can be used to enlarge the set of confined $p$-form fields. With this, we will generalize that study performed by \cite{Duff} in codimension one. Next, we discuss the consistency of the localization of these fields with Einstein's equations in this new gravitational configuration. Therefore, this manuscript intends to reinforce that the finite integral argument is a necessary, but not sufficient, condition to provide a consistent effective field theory over the brane. And also, that some symmetries of the theory can be used to include other fields into the set of confined fields. This work is organized as follows: in section (\ref{Sec-2}), we will make a brief review of some results presented in the literature about $p$-form confinement for some specific codimension two braneworld models. In section (\ref{Sec-3}), we will explore the Hodge duality and its consequences on the fields' localization. Finally, we will discuss the consistency of the localization with the Einstein's equations in section (\ref{Sec-4}). The conclusions are left to the section (\ref{Sec-5}).

\section{$p$-Form Fields Localization - Review}\label{Sec-2}
In this section we will review some results about localization of $p$-form fields on codimension two braneworlds found in the literature. In doing this, we first briefly describe the background metric considered in each case. Later, this will be used to verify the generic results which will be discussed in next sections.

\begin{enumerate}
\item[\hypertarget{(2A)}{(2A)}] First, let us consider the string-like braneworld presented in Refs. \cite{Oda02, Oda03}, which is a generalization of others presented early in Refs. \cite{Gherghetta02, Gherghetta}. This brane model is obtained from the action
\begin{equation}\label{EQ-2A-01}
S_{\mbox{grav.}}=\int d^{d}xdrd\theta\sqrt{-g}\left[\frac{1}{2\kappa^{2}}\left(R-2\Lambda\right)+\mathcal{L}_{b}\right],
\end{equation}
where the energy-momentum tensor is given by $T_{M}^{(b)N}=\left(\delta_{\mu}^{\nu}t_{0}(r),t_{r}(r),t_{\theta}(r)\right)$. The background metric for this case is written as
\begin{eqnarray}\label{EQ-2A-02}
ds^{2}=e^{-2kr}\eta_{\mu\nu}dx^{\mu}dx^{\nu}+e^{-2B(r)}d\theta^{2}+dr^{2}.
\end{eqnarray}
In the above equation, $\eta_{\mu\nu}$ is the Minkowski metric on the brane and $k$ is a positive constant defined as
\begin{eqnarray}\label{EQ-2A-03}
k^{2}=\frac{2\kappa^{2}_{D}t_{\theta}-2\Lambda}{d(d+1)}>0.
\end{eqnarray}
The warp factor $B(r)$ is obtained from Einstein's equation and it is given by
\begin{eqnarray}\label{EQ-2A-04}
B(r)=kr+\frac{2\kappa^{2}_{D}}{kd}\int^{r}dr'(t_{r}-t_{\theta}).
\end{eqnarray}
From this, we can discuss two particular solutions for the metric. Namely, the global defect case ($t_{r}=-t_{\theta}=constant$), where $B(r)$ is given by
$$B(r)=\left[k-2\frac{\kappa^{2}_{D}t_{\theta}}{kd}\right]r\equiv(k-\delta)r$$
and the local defect case where $B(r)=kr$. This last one can be obtained as a particular case of the above result by putting $\delta=0$ (without sources). By choosing $t_{r}=t_{\theta}$ in (\ref{EQ-2A-04}) we obtain the other case discussed in \cite{Gherghetta02, Gherghetta}. For the below discussion, we will consider only the global defect case and that without sources. As discussed in \cite{Oda03}, the localization of gravity in this model is obtained for 
\begin{equation}\label{EQ-2A-05}
-\frac{|\Lambda|}{\kappa^{2}_{D}}<t_{\theta}<\frac{(d-1)|\Lambda|}{2\kappa^{2}_{D}},
\end{equation}
where we made explicit the negative sign of the cosmological constant. 

Now, with the background defined, we can review the $p$-form field localization.
The confinement of the free scalar ($0$-form) and vector ($1$-form) fields in this scenario can be found in Refs. \cite{Oda01, Oda02, Oda03}. After, Ref. \cite{Alencar07} generalized this study for a free $p$-form field. To include these cases in a single approach, we will discuss below the results of Ref. \cite{Alencar07} for an arbitrary $p$-form field $\mathcal{A}_{N_{1}...N_{p}}$. The authors started from an action given by
\begin{eqnarray}\label{EQ-2A-06}
S^{(\mbox{mat})}=-\frac{1}{2(p+1)!}\int d^{D}x \sqrt{-g}\mathcal{F}_{N_{1}...N_{p+1}}\mathcal{F}^{N_{1}...N_{p+1}}.
\end{eqnarray}
In this action, $\mathcal{F}_{N_{1}...N_{p+1}}=\partial_{[N_{1}}\mathcal{A}_{N_{2}...N_{p+1}]}$, $g$ is the determinant of the metric (\ref{EQ-2A-02}) and $D=d+2$, where $d$ is the brane dimension. Throughout the manuscript, capital indexes $M,N$ run on all dimensions $D=d+2$. Coordinates $x^{\mu}$ span the brane and $\mu,\nu=(1,2,..,d)$. The next mathematical steps are similar for most of the above references. From equation (\ref{EQ-2A-06}), the authors get the equation of motion
\begin{eqnarray}\label{EQ-2A-07}
\partial_{N_{1}}\left[\sqrt{-g}\mathcal{F}^{N_{1}...N_{p+1}}\right]=0.
\end{eqnarray}
From this, by considering only the components $\mathcal{A}_{\mu_{1}...\mu_{p}}$ nonzero and using the metric (\ref{EQ-2A-02}), they get
\begin{eqnarray}\label{EQ-2A-08}
\partial_{\mu_{1}}\mathcal{F}^{\mu_{1}\mu_{2}...\mu_{p+1}}&+&e^{[d-2(p+1)]kr+B(r)}\partial_{r}\left[e^{-[d-2p]kr-B(r)}\partial_{r}\mathcal{A}^{\mu_{2}...\mu_{p+1}}\right]\nonumber\\&+&e^{[d-2(p+1)]kr+B(r)}\partial_{\theta}\left[e^{-[d-2(p+1)]kr+B(r)}g^{\theta\theta}\partial_{\theta}\mathcal{A}^{\mu_{2}...\mu_{p+1}}\right]=0.
\end{eqnarray}
Where the index contractions are performed with the Minkowski metric. Next, by proposing the variable separation $\mathcal{A}_{\mu_{1}...\mu_{p}}(x,r,\theta)=A_{\mu_{1}...\mu_{p}}(x)\xi(r)$ for the s-state \footnote{Here, s-state means that the solution does not depend on the $\theta$ coordinate.}, they get
\begin{eqnarray}
\partial_{\mu_{1}}F^{\mu_{1}\mu_{2}...\mu_{p+1}}(x)&=&m^2A^{\mu_{2}...\mu_{p+1}}(x),\\\label{EQ-2A-09}
e^{[d-2(p+1)]kr+B(r)}\partial_{r}\left[e^{-[d-2p]kr-B(r)}\partial_{r}\xi(r)\right]&=&m^2\xi(r).\label{EQ-2A-10}
\end{eqnarray}
This was obtained for a $p$-form field (s-state) in Ref. \cite{Alencar07}, for the scalar field in \cite{Oda03} and for the vector field in \cite{Oda01, Oda03}. From these results, we can solve the equation (\ref{EQ-2A-10}) for the zero-mode ($m^{2}=0$) and, with this, discuss the localization in the action (\ref{EQ-2A-06}).

Following the procedure of the above references, equation (\ref{EQ-2A-10}) have a constant solution for the zero-mode ($m^{2}=0$). With this, they get a confined field when the integral
\begin{eqnarray}\label{EQ-2A-11}
K=R_{0}\int drd\theta e^{-[d-2p-1]kr+\delta r}\xi_{0}^{2}(r)=R_{0}c^{2}_{1}\int drd\theta e^{-[d-2p-1]kr+\delta r},
\end{eqnarray}
obtained from the action (\ref{EQ-2A-06}), is finite. Finally, from this, the localization condition is obtained, i.e., the integral $K$ is finite, for 
$$
(d-2p-1)k>\delta.
$$ 
For the local defect, where $\delta=0$ [$t_{\theta}=0$], it is easy to obtain 
\begin{equation} \label{localdefectcondition}
p<\frac{(d-1)}{2}.
\end{equation}
Therefore, if $d=4$, only the scalar ($p=0$) and the vector ($p=1$) fields can be confined by using the constant solution $\xi_{0}$.
For the global defect, the above condition must be supplemented with the condition for gravity localization (\ref{EQ-2A-05}). Thus, these two constraints together give us 
\begin{equation}
 -\frac{|\Lambda|}{\kappa^{2}_{D}}<t_{\theta}<\frac{(3-2p)|\Lambda|}{2(1+p)\kappa^{2}_{D}}.\label{EQ-2A-12}
\end{equation}
In this case, any $p$-form field can be confined by a suitable choice of the parameter $t_{\theta}$. This is due to the fact that  
$$
-\frac{|\Lambda|}{\kappa^{2}_{D}}<\frac{(3-2p)|\Lambda|}{2(1+p)\kappa^{2}_{D}}\leq\frac{(d-1)|\Lambda|}{2\kappa^{2}_{D}}
$$
for any value of $p$. These results, for $t_{\theta}=0$ or $t_{\theta}\neq0$, has been found in the Refs. \cite{Oda01, Oda03} for the scalar and the vector fields, and, after, generalized for any $p$-form field in Ref. \cite{Alencar07}.


\item[\hypertarget{(2B)}{(2B)}] Now, let us consider the brane model presented in Ref. \cite{Silva}. This model describes the spacetime inside and outside a `thick' string-like topological defect with an asymptotically AdS$_{6}$ spacetime. Just like the previous case, the extra dimension are $r\in[0,\infty)$ and $\theta\in[0,2\pi)$, and the action is similar to that in equation (\ref{EQ-2A-01}). However, for this braneworld, the metric is given by
\begin{eqnarray}\label{EQ-2B-01}
ds^{2}=e^{-kr+\tanh(kr)}\left[dx^{\mu}dx_{\mu}+\tanh^{2}(kr)k^{-2}d\theta^{2}\right]+dr^{2},
\end{eqnarray}
where $k$ is a positive constant related to the cosmological constant. As discussed in Ref. \cite{Silva}, for $r\to\infty$, the above metric gives the solution discussed in Refs. \cite{Gherghetta02, Gherghetta}, which is a particular solution of the previous case \cite{Oda01, Oda02, Oda03}. However, the model (\ref{EQ-2B-01}) has the advantage of being valid inside the string-like core. The Ricci scalar for this metric, by considering a flat brane, is given by 
\begin{eqnarray}\label{EQ-2B-02}
R=-k^{2}\left[\frac{15}{2}\tanh^{4}(kr)-16\sech^{2}(kr)\tanh(kr)-4\sech^{2}(kr)\right].
\end{eqnarray}
As we can see, this function is completely regular for all $r$ and, for $r\to\infty$, it gets the constant value
$$ R_{\infty}=-\frac{15}{2}k^{2}.$$
Therefore this is an asymptotically  AdS$_{6}$ spacetime. We will end the characterization of the background here, for more details see the Ref. \cite{Silva}.

Now, let us review the vector field localization in the above background. This was performed in Ref. \cite{Costa02}.The authors start from an action given by
\begin{eqnarray}\label{EQ-2B-03}
S^{(\mbox{mat})}=-\frac{1}{4}\int d^{6}x\sqrt{-g}\mathcal{F}_{NM}\mathcal{F}^{NM},
\end{eqnarray}
with $\mathcal{F}_{MN}=\partial_{M}\mathcal{A}_{N}-\partial_{N}\mathcal{A}_{M}$. The steps are similar to those presented in the previous case. From the action (\ref{EQ-2B-03}), they obtain the equation of motion
\begin{eqnarray}
\partial_{N}\left[\sqrt{-g}\mathcal{F}^{NM}\right]=0.\label{EQ-2B-04}
\end{eqnarray}
In order to solve the above equation the authors choose the gauge $\partial_{\mu}\mathcal{A}^{\mu}=0$ and also the particular solutions $\mathcal{A}_{r}$ and $\mathcal{A}_{\theta}$ constants. They get an equation similar to (\ref{EQ-2A-08}) for $p=1$. Next, by proposing the separation of variable $\mathcal{A}_{\mu}(x,r)=A_{\mu}(x)\xi(r)$, they get for the zero-mode (s-state)
\begin{eqnarray}
\partial_{\mu}F^{\mu\nu}(x)=0,\\ \label{EQ-2B-05}
\partial_{r}\left[e^{\frac{3}{2}[-kr+\tanh(kr)]}\tanh(kr)\partial_{r}\xi_{0}(r)\right]=0.\label{EQ-2B-06}
\end{eqnarray}
From this, unlike the previous case, the authors obtained two zero-mode (s-state) solutions from (\ref{EQ-2B-06}), which are given by
\begin{eqnarray}
\xi_{0,(1)}(r)=c_{1}, \hspace{1cm} \xi_{0,(2)}(r)=c_{2}\int^{r} \frac{e^{-\frac{3}{2}\left[-kr'+\tanh(kr')\right]}}{\tanh(kr')}dr'.\label{EQ-2B-07}
\end{eqnarray}
With this they get  
\begin{eqnarray}
K\propto \int dr e^{\frac{1}{2}\left[-kr+\tanh(kr)\right]}\tanh(kr)\xi^{2}_{0}(r),\label{EQ-2B-08}
\end{eqnarray}
which must be analyzed with both solutions (\ref{EQ-2B-07}). To see this we will need of the asymptotic behavior of the integrands. 

Let us analyze the constant solution $\xi_{0,(1)}(r)$. With this, the behavior of the integrand of $K$ is regular for all values of $r$, thus, its convergence is determined in the limit of $r\to\infty$. For large $r$ we have that
$$
e^{\frac{1}{2}\left[-kr+\tanh(kr)\right]}\tanh(kr)\xi^{2}_{0,(1)}(r)\to e^{-\frac{1}{2}kr}
$$
and therefore the zero-mode is localized with the constant solution.  
 
The other possibility for the integral $K$ is to use the non-constant solution $\xi_{0,(2)}(r)$, but, in doing this, we must be more careful. To start, let us make some comments about this solution. As we can observe from equation (\ref{EQ-2B-07}), the function at that integral is singular for $r\to0$ and, in this limit, it gives 
$$
\xi_{p,0,(2)}(r\to 0)\propto \ln(kr).
$$ 
However the integrand of  $K$ is regular for $r\to 0$ since 
$$
e^{\frac{1}{2}[-kr+\tanh(kr)]}\tanh(kr)\xi^{2}_{0,(2)}(r)\to kr[\ln(kr)]^{2}\to 0.
$$
Therefore, although the solution $\xi_{0,(2)}(r)$ is singular, the function at the integral $K$ is regular in this limit. Thus, the convergence of the complete integral (\ref{EQ-2B-07}) with $\xi_{0,(2)}(r)$ is determined by its behavior in the limit $r\to \infty$. In this limit, for $\xi_{0,(2)}(r)$ we have
$$\xi_{0,(2)}(r)(r\to\infty)\propto e^{\frac{3}{2}kr}.$$
Therefore the asymptotic behavior of our integrand will be given by 
$$
e^{\frac{1}{2}\left[-kr+\tanh(kr)\right]}\tanh(kr)\xi^{2}_{0,(2)}(r)\to e^{\frac{5}{2}kr},
$$
which is not normalizable. Therefore this solution must be discarded for the vector field.

Therefore, for this model, the vector field is confined  only with the constant solution $\xi_{0,(1)}(r)=c_1$ \cite{Costa02}. This is important since, when we generalize the above results do $p$-forms, the non-constant solution will be used to localize some effective fields.


\item[\hypertarget{(2C)}{(2C)}] As a third example, we will discuss the braneworld model presented in Ref. \cite{Costa2013}. In this reference, the authors build a $6$D spacetime in the form $\mathcal{M}_{6}=\mathcal{M}_{4}\times\mathcal{C}_{2}$, where $\mathcal{C}_{2}$ is a two-cycle of the resolved conifold \cite{Zayas} and $\mathcal{M}_{4}$ is the $3$-brane. Just like the previous cases, the extra dimensions are in the range $r\in[0,\infty)$ and $\theta\in[0,2\pi)$. The metric for this scenario is given by
\begin{eqnarray}\label{EQ-2C-01}
ds^{2}=e^{-kr+\tanh(kr)}\left[dx^{\mu}dx_{\mu}+\beta^{2}(a,r)d\theta^{2}\right]+dr^{2}.
\end{eqnarray}
In the above metric, $k$ is a positive constant and the parameter $a$ measure how smooth is the conifold (conical singularity in $r=0$). The function $\beta(a,r)$ is defined as
\begin{eqnarray}\label{EQ-2C-02}
\beta^{2}(a,r)=a^{2}+u^{2}(a,r),\ \ \ \ \ u(a,r)=\frac{1}{\sqrt{6}}\left\lbrace\begin{array}{r} r\hspace{2cm} , a=0; \\ -i\sqrt{6}aE\!\left(\mbox{arcsinh}\!\left(\frac{i}{\sqrt{6}a}r\right),\frac{2}{3}\right), a\neq 0.\end{array}\right.
\end{eqnarray}
The function $E$ is the elliptic integral of second kind. As showed in the Ref. \cite{Costa2013}, the Ricci scalar is regular for all $r$ when $a\neq0$, and it goes to a negative constant value for $r\to\infty$. Therefore, the metric (\ref{EQ-2C-01}) presents an asymptotically AdS$_{6}$ characteristic.

Moreover, to describe the features the spacetime background, the same authors study the gauge field localization in this scenario. To do this, they start from standard gauge field action (\ref{EQ-2B-03}). After following the same steps as in the case \hyperlink{(2B)}{(2B)} , they get for the zero-mode (s-state) the equations
\begin{eqnarray}
\partial_{\mu}F^{\mu\nu}(x)=0,\\ 
\partial_{r}\left[e^{\frac{3}{2}[-kr+\tanh(kr)]}\beta(a,r)\partial_{r}\xi_{0}(r)\right]=0.\label{EQ-2C-04}
\end{eqnarray}
Where $\xi_{0}(r)$ comes from the separation of variable $\mathcal{A}_{\mu}(x,r)=A_{\mu}(x)\xi_0 (r)$. From this equation of motion, the author find the solution
\begin{eqnarray}
\xi_{0,(1)}(r)=c_{1}.\label{EQ-2C-05}
\end{eqnarray}
With this they obtain 
\begin{eqnarray}
K\propto \int dr e^{\frac{1}{2}\left[-kr+\tanh(kr)\right]}\beta(a,r)\xi^{2}_{0}(r),\label{EQ-2C-06}
\end{eqnarray}
which will be finite because the transverse space, spanned by $(r,\theta)$, has a finite volume. Therefore, the vector field is confined for this model with the constant solution (\ref{EQ-2C-05}) \cite{Costa2013}. Just like the previous cases, we will discuss the consistency of the above results in next sections. Beyond this, this study will also be generalized to include the free $p$-form field.

\end{enumerate}

This review gives us some intuition about fields localization in different braneworld models. These models will be used to discuss the generic results presented in next sections for the $p$-form fields. In doing this, let us perform the discussion about the consistency of the above results. Next, we will show how the Hodge duality can provide the consistent localization of these fields.

\section{Consistency Conditions: Hodge duality}\label{Sec-3}

From now on, we will discuss the confinement of a free massless $p$-form field in a generic codimension two braneworld. In this context, we will explore the Hodge duality (HD) and its consequences on the localization of these fields. The review performed in the previous section does not consider this symmetry and, as we will see, it has some interesting consequences. This symmetry has been considered in Ref. \cite{Duff}, where the authors studied the consistency of the $p$-form localization with the Hodge duality and also with the Einstein's equations (EE) for codimension one models. The authors showed that the $p$-form must satisfy not only the finite integral requirement, but also other constraints obtained from HD and EE. In this section, we will discuss the Hodge duality and how it can be used to enlarge the set of $p$-form fields confined. Thus, we will generalize the results of the Ref. \cite{Duff} for the codimension two scenario. However, before to discuss this new dimensional configuration, we will make a brief review of the codimension one case.

\subsection{Codimension one case}\label{Subsec-3-1}

In Ref. \cite{Duff}, the authors discuss the consistency of the results presented previously in the literature about the localization of the $p$-form field. According to them, such results are not in accordance with the Hodge duality. Due to the importance for us, below we describe the essence of the results presented in Ref. \cite{Duff}. 

The authors consider a codimension one background with metric given by
\begin{equation}\label{EQ-3-1-01}
ds^{2}=e^{-2k|z|}\hat{g}_{\mu\nu}(x)dx^{\mu}dx^{\nu}+dz^{2}.
\end{equation}
In this context, the action for a free massless bulk $p$-form field $\mathcal{A}_{M_{1}...M_{p}}$ is written as
\begin{equation}\label{EQ-3-1-02}
S=-\frac{1}{2(p+1)!}\int d^{d}xdz\sqrt{-g}\mathcal{F}_{M_{1}...M_{p+1}}\mathcal{F}^{M_{1}...M_{p+1}}.
\end{equation}
Where $\mathcal{F}_{M_{1}...M_{p+1}}=\partial_{[M_{1}}\mathcal{A}_{M_{2}...M_{p+1}]}$, $d$ is the brane dimension and $z$ is the extra dimension. The equations of motion obtained from the above action are
\begin{equation}\label{EQ-3-1-03}
\partial_{M_{1}}\left[\sqrt{-g}\mathcal{F}^{M_{1}...M_{p+1}}\right]=0.
\end{equation}
From this, we can follow the common procedure. For codimension one case, the $p$-form $\mathcal{A}_{M_{1}...M_{p}}$ has the components $\mathcal{A}_{\mu_{1}...\mu_{p}}$ and $\mathcal{A}_{\mu_{1}...\mu_{p-1}z}$. However, due to the gauge invariance of the action, the components $\mathcal{A}_{\mu_{1}...\mu_{p-1}z}$ can be eliminated. Therefore, we can consider only the components $\mathcal{A}_{\mu_{1}...\mu_{p}}$ nonzero. Next, we perform the separation of variable $\mathcal{A}_{\mu_{1}...\mu_{p}}(x,z)=A_{\mu_{1}...\mu_{p}}(x)\xi(z)$ to get the equations for the zero-mode given by
\begin{eqnarray}
\partial_{\mu_{1}}\left[\sqrt{-\hat{g}(x)}F^{\mu_{1}...\mu_{p+1}}(x)\right]=0,\label{EQ-3-1-04}\\
\partial_{z}\left[e^{-(d-2p)k|z|}\partial_{z}\xi_{0}(z)\right]=0.\label{EQ-3-1-05}
\end{eqnarray}
Here, we already used the metric (\ref{EQ-3-1-01}). Equation (\ref{EQ-3-1-05}) has a constant solution and, with this, the action (\ref{EQ-3-1-02}) can be written for the zero-mode as
\begin{eqnarray}\label{EQ-3-1-06}
S_{0}=-\frac{1}{2(p+1)!}\int d^{d}x\sqrt{-\hat{g}(x)}\hat{F}_{\mu_{1}...\mu_{p+1}}(x)\hat{F}^{\mu_{1}...\mu_{p+1}}(x)\int dz e^{-\left[d-2(p+1)\right]k|z|}.
\end{eqnarray}
In this way, the $p$-form fields confined with the constant solution are those where
\begin{equation}\label{EQ-3-1-07}
p<\frac{d-2}{2}.
\end{equation}
Thus, for the particular case of a $3$-brane only the free massless $0$-form can be localized. This is the result contested in Ref. \cite{Duff}.

The authors argue that the above results are in contradiction with the Hodge duality. As discussed in Ref. \cite{Duff80}, in the absence of topological obstructions, a free $p$-form $A_{[p]}$ in the bulk is dual to a free $(d-p-1)$-form $B_{[d-p-1]}$ with field strength $(\star F)_{[d-p]}={\bf d} B_{[d-p-1]}$. We will use the notation [$\mathcal{A}_{M_{1}...M_{p}}$, $\mathcal{B}_{M_{1}...M_{d-p-1}}$] for the dual fields. Beyond this, according to the authors, to use the constant solution for $A_{[p]}$ and $B_{[d-p-1]}$ is not compatible with the Hodge duality transformation 
\begin{equation}\label{EQ-3-1-08}
(\star F)^{M_{1}...M_{d-p}}=\frac{1}{(p+1)!\sqrt{-g}}\epsilon^{M_{1}...M_{d-p}N_{1}...N_{p+1}}F_{N_{1}...N_{p+1}}.
\end{equation}
Thus, the result (\ref{EQ-3-1-07}) is not consistent because, in $4$D, for example, the scalar field is dual to a $2$-form, and this last one is not allowed by the above results. To solve this and to include other $p$-form with higher $p$, they simply write the other solution of equation (\ref{EQ-3-1-05}), which is given by
\begin{eqnarray}\label{EQ-3-1-10}
\xi_{0,(2)}(z)=c_{2}e^{(d-2p)k|z|}.
\end{eqnarray}
In this way, they use the constant solution to confine the $p$-form with $p<(d-2)/2$ and the solution (\ref{EQ-3-1-10}) for the other cases. With this, the Hodge duality (\ref{EQ-3-1-08}) can be written as
\begin{eqnarray}
(\star \mathcal{F})^{\nu_{1}...\nu_{d-p-1}z}&=&\frac{1}{(p+1)!\sqrt{-g}}\epsilon^{\nu_{1}...\nu_{d-p-1}z\mu_{1}...\mu_{p+1}}\mathcal{F}_{\mu_{1}...\mu_{p+1}},\label{EQ-3-1-11}\\
(\star \mathcal{F})^{\nu_{1}...\nu_{d-p}}&=&\partial_{[\nu_{1}}\mathcal{B}_{\nu_{2}...\nu_{d-p}]}=0.\label{EQ-3-1-12}
\end{eqnarray}
From the equation (\ref{EQ-3-1-12}), we get that $\mathcal{B}_{\mu_{1}...\mu_{d-p-1}}=\partial_{[\mu_{1}}\mathcal{C}_{\mu_{2}...\mu_{d-p-1}]}$ and, from (\ref{EQ-3-1-11}), we get
\begin{eqnarray}
B^{\nu_{1}...\nu_{d-p-1}}(x)=\frac{1}{(p+1)!\sqrt{-\hat{g}(x)}}\epsilon^{\nu_{1}...\nu_{d-p-1}z\mu_{1}...\mu_{p+1}}F_{\mu_{1}...\mu_{p+1}}(x).\label{EQ-3-1-13}
\end{eqnarray}
Thus, the duality in the bulk, described by (\ref{EQ-3-1-08}), is preserved on the brane, as showed by the above relation. However, all this is consistent only if both fields can be confined. Therefore, we need verify if the components $B^{\nu_{1}...\nu_{p'}}(x)$ are confined with (\ref{EQ-3-1-10}).

With the solution (\ref{EQ-3-1-10}), the action (\ref{EQ-3-1-02}) can be written as
\begin{equation}\label{EQ-3-1-14}
S=-\frac{1}{2p'!}\int d^{d}x\sqrt{-\hat{g}(x)}B_{\mu_{1}...\mu_{p'}}(x)B^{\mu_{1}...\mu_{p'}}(x)\int dz e^{(d-2p')k|z|}.
\end{equation}
And the localization is attained for $p'>d/2$. Note that the bulk $p$-form fields confined with the solution  (\ref{EQ-3-1-10}) are found on the brane as $(p'-1)$-form, namely, the field $\mathcal{C}_{\mu_{1}...\mu_{p'-1}}$. By considering $p'=d-p-1$, the effective field on the $(d-1)$-brane is a $(d-p-2)$-form which is exactly the Hodge dual of the $p$-form, preserving, thus, the bulk Hodge duality on the $(d-1)$-brane.

\begin{figure}
\centering
\includegraphics[scale=0.8,trim = 4.3cm 13.3cm 3.5cm 13.3cm,clip]{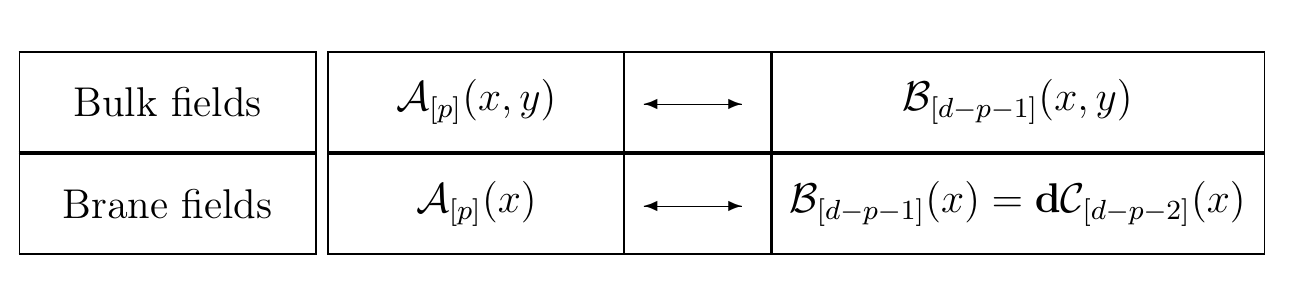}
\caption{Equivalency relation (indicated by the arrows) between the bulk fields, and also, between the effective brane fields after the dimensional reduction.}\label{Figura-00}
\end{figure}

The figure (\ref{Figura-00}) shows the equivalence relation between the bulk dual fields and also between the effective fields. For example, the $0$-form is bulk dual to a $3$-form in $5$D, and these fields are confined on a $3$-brane as a $0$-form and a $2$-form, which are dual in $4$D. Thus, the Hodge duality provides us with others confined fields for this codimension one scenario. These are the main results presented in \cite{Duff}. Below, we will generalize this discussion for the codimension two case.

\subsection{Codimension two case}\label{Subsec-3-2}

As we saw above, Ref. \cite{Duff} explored a symmetry of the theory to drive a correct description of the localization. To discuss this, let us start by writing the Lagrangian for a free massless $p$-form field
\begin{equation}\label{EQ-3-2-01}
\mathcal{L}=-\frac{1}{2(p+1)!}\mathcal{F}_{M_{1}...M_{p+1}}\mathcal{F}^{M_{1}...M_{p+1}},
\end{equation}
where $\mathcal{F}_{M_{1}...M_{p+1}}$ are the components of $\mathcal{F}_{[p+1]}=\ ${\bf d}$\mathcal{A}_{[p]}$. The operator `{\bf d}' is the exterior derivative. The Lagrangian (\ref{EQ-3-2-01}) presents a gauge symmetry provided by the transformation
\begin{eqnarray}
\mathcal{A}'_{[p]}=\mathcal{A}_{[p]}+\mbox{\bf d}\theta_{[p-1]}.
\end{eqnarray}
With this and by using the fact that the $p$-form is massless, we can show that the field $\mathcal{A}_{[p]}$ has 
$$_{D-2}C_{p}\equiv\left(\begin{array}{c}
D-2\\p
\end{array}\right)$$
degree of freedom after the complete gauge fixing. Another well-known fact is that the Hodge duality (HD) transformation relates a $q$-form to a $(D-q)$-form, where $D$ is the spacetime dimension. In Ref. \cite{Duff80}, the authors showed that, by considering the Hodge duality in a spacetime without topological obstructions, there must be an equivalence among a $p$-form field $\mathcal{A}_{[p]}$ and a $(D-p-2)$-form field $\mathcal{B}_{[D-p-2]}$. Below, we show this equivalence.

Let us consider the Hodge duality transformation
\begin{equation}\label{EQ-3-2-02}
\left(\star \mathcal{F}\right)^{M_{1}...M_{D-p-1}}=-\frac{(-1)^{(p+1)(D-p-1)}}{(p+1)!\sqrt{-g}}\varepsilon^{M_{1}...M_{D-p-1}N_{1}...N_{p+1}}\mathcal{F}_{N_{1}...N_{p+1}},
\end{equation}
where $\left(\star\mathcal{F}\right)_{M_{1}...M_{D-p-1}}$ are the components of $(\star\mathcal{F})_{[D-p-1]}\equiv\ ${\bf d}$\mathcal{B}_{[D-p-2]}$. With this, it is easy to obtain that 
\begin{eqnarray}\label{EQ-3-2-03}
\mathcal{L}=-\frac{1}{(p\!+\!1)!}\mathcal{F}_{M_{1}..M_{p+1}}\mathcal{F}^{M_{1}..M_{p+1}}=-\frac{1}{(D\!-\!p\!-\!1)!}\left(\star \mathcal{F}\right)^{N_{1}..N_{D-p-1}}\left(\star \mathcal{F}\right)_{N_{1}..N_{D-p-1}}.
\end{eqnarray}
In this sense, we get a {\it naive equivalence}, provided by the Hodge duality transformation, between the free massless bulk fields $\mathcal{A}_{[p]}$ and $\mathcal{B}_{[D-p-2]}$. Note that the HD transformation hold for the field strengths independently the equations of motion, therefore it holds off-shell for the field strengths. However, the physical degrees of freedom are contained in the fields $p$-form and $(D-p-2)$-form. As we discussed above, $p$-form has $_{D-2}C_{p}$ degrees of freedom (on-shell). Since $_{D-2}C_{p}=_{D-2}C_{D-p-2}$ and $_{D-2}C_{D-p-2}$ is the number of independent components of a $(D-p-2)$-form after the complete gauge fixing, the $p$-form and the $(D-p-2)$-form have the same degree of freedom (on-shell) and, therefore, they are really equivalent \cite{Tanii}. To conclude, even if the Hodge duality transformation is applied to the field strengths, the fields $p$-form and $(D-p-2)$-form, which contain the physical degrees of freedom and the same number of them on-shell, will be called the dual fields. Based on this idea, the authors in \cite{Duff} showed that the localization of a free massless $p$-form must manifest this equivalence for codimension one braneworlds. Below, we will generalize this for codimension two braneworld models.

The above equivalence has important consequences on the study of the free $p$-form field localization. 
These consequences can be expressed by the following statements:
\begin{itemize}
\item[-] {\bf Statement (i):} {\it Confinement must be possible for both fields, the bulk $p$-form and its bulk dual $(D-p-2)$-form. In other words, localization of a free massless bulk $p$-form must imply the localization of its bulk dual $(D-p-2)$-form.}\\
This is an immediate consequence of equation (\ref{EQ-3-2-03}). As the Hodge duality transformation is a symmetry of theory, the action obtained for (\ref{EQ-3-2-03}) should be the same for both dual fields. This is possible only if the integral in extra dimensions is finite for both fields, for the $p$-form and also for its bulk dual $(D-p-2)$-form.

\item[-] {\bf Statement (ii):} {\it Hodge duality must be valid even after the dimensional reduction for the effective fields on the brane.}
\\
This is an immediate consequence of (\ref{EQ-3-2-02}). 
\end{itemize}
The above statements are simple, but they give us important information about the confined $p$-form fields.  Before considering specific cases in section (\ref{Sec-4}), we can already obtain some direct consequences about the effective fields in general codimension two braneworlds.

The main point is that there is a crucial difference between the codimension one and two cases. In both cases, the effective field $\mathcal{A}_{\mu_{1}...\mu_{p-1}z}$ can be eliminated by using gauge symmetry. For the codimension one case we are left only with the component $\mathcal{A}_{\mu_{1}...\mu_{p}}$ and the analyses becomes simplified.  However, in the codimension two case, we also have the effective fields 
$$
\mathcal{A}_{\mu_{1}...\mu_{p-1}w},\mathcal{A}_{\mu_{1}...\mu_{p-2}zw}.
$$ 
In the above equation we are using $w,z$ for the extra dimensions, which can be $r,\theta$ for example. The general procedure found in the literature (see \cite{Alencar07}) is to consider the particular solution 
$$
\mathcal{A}_{\mu_{1}...\mu_{p-1}w}=\mathcal{A}_{\mu_{1}...\mu_{p-2}wz}=0.
$$
However this is not consistent with the HD and therefore with the statements above.  The reason is that, in codimension two case, the equivalence in the bulk is between a $p$-form and a $(d-p)$-form. By using this in the statement (ii), we must have a $p$-form and its brane dual $(d-p-2)$-form localized over the $(d-1)$-brane. The brane dual will come exactly from the components $\mathcal{A}_{\mu_{1}...\mu_{p-2}lm}$ and therefore are crucial to keep the Hodge duality and, with this, the consistency of the model. Below we will use this fact to discuss the localization of a $p$-form for a generic codimension two scenario and its consistency with the Hodge duality. After this, we will apply our results to the models of the section (\ref{Sec-2}).

\subsection{$p$-form fields localization}\label{Subsec-3-3}

The results presented in section \ref{Sec-2} do not consider the Hodge duality. Unlike the codimension one case, it is not possible to solve this problem only with the components $\mathcal{A}_{\mu_{1}...\mu_{p}}$. As said above, the  $\mathcal{A}_{\mu_{1}...\mu_{p-2}lm}$ must be present in order to preserve Hodge duality. Since nowhere in the literature this is done, we develop below a more complete description of the $p$-form field localization.

Let us start by considering an arbitrary codimension two braneworld background given by
\begin{equation}\label{EQ-3-3-01}
ds^{2}=g_{MN}dx^{N}dx^{M}=e^{2\sigma(y)}\hat{g}_{\mu\nu}(x)dx^{\mu}dx^{\nu}+\bar{g}_{jk}(y)dy^{j}dy^{k},
\end{equation}
where the warp factor $\sigma(y)$ and the metric components $\bar{g}_{jk}(y)$ depend on the extra dimensions coordinates $y^{j}$. Capital indexes $M,N$ run on all dimensions $D=d+2$. Coordinates $x^{\mu}$ span the brane and $\mu,\nu=(1,2,..,d)$ and coordinates $y^{j}$ are related to the extra dimensions with $j,k=(1,2)$. In this scenario, the metric (\ref{EQ-3-3-01}) is completely generic and, at first, we will not need to know it. However, let us assume that the gravitational action which gives the background solution (\ref{EQ-3-3-01}) is  
\begin{equation}\label{EQ-3-3-02}
 S_{grav.}=\int d^{d}xd^{2}y \sqrt{-g}\left[\frac{1}{2\kappa}\left(R -2\Lambda\right)+ \mathcal{L}_{b}\left(y\right)\right],
 \end{equation} 
where $\Lambda$ is the cosmological constant and $\mathcal{L}_{b}\left(y\right)$ is the matter source of the brane. In this context, the metric (\ref{EQ-3-3-01}) will be called vacuum solution.

With the background metric solution previously defined, let us start the study of a free $p$-form field $\mathcal{A}_{N_{1}...N_{p}}$ localization in such background. The action for this field is given by
\begin{equation}\label{EQ-3-3-04}
S=-\frac{1}{2(p+1)!}\int d^{d}xd^{2}y\sqrt{-g}\mathcal{F}_{N_{1}...N_{p+1}}\mathcal{F}^{N_{1}...N_{p+1}},
\end{equation}
where $\mathcal{F}_{N_{1}..N_{p+1}}=\partial_{[N_{1}}\mathcal{A}_{N_{2}..N_{p+1}]}$ and $g$ is the determinant of the metric (\ref{EQ-3-3-01}). From this action, the equations of motion (EOM) can be written as
\begin{eqnarray}\label{EQ-3-3-05}
\partial_{\mu_{1}}\!\!\left[\sqrt{-g}\mathcal{F}^{\mu_{1}N_{2}...N_{p+1}}\right]+\partial_{k}\!\!\left[\sqrt{-g}\mathcal{F}^{kN_{2}...N_{p+1}}\right]=0.
\end{eqnarray}
Now, by using the metric (\ref{EQ-3-3-01}), we get 
\begin{eqnarray}
\frac{1}{\sqrt{-\hat{g}}}\partial_{\rho}\!\left[\sqrt{-\hat{g}}\mathcal{F}^{\rho\mu_{1}..\mu_{p}}\right]+\frac{e^{2\sigma}}{\tilde{H}(y)}\partial_{k}\!\left[\tilde{H}(y)\bar{g}^{kj}\left(\partial_{j}\mathcal{A}^{\mu_{1}..\mu_{p}}+(-1)^{p}\partial^{[\mu_{1}}\mathcal{A}^{\mu_{2}..\mu_{p}]}_{\ \ \ \ \ \ \ j}\right)\right]=0,\label{EQ-3-3-06}\\
\frac{1}{\sqrt{-\hat{g}}}\partial_{\rho}\!\left[\sqrt{-\hat{g}}\left(\partial_{m}\mathcal{A}^{\rho\mu_{1}..\mu_{p-1}}+(-1)^{p}\partial^{[\rho}\mathcal{A}^{\mu_{1}..\mu_{p-1}]}_{\ \ \ \ \ \ \ \ \ \ m}\right)\right]\ \ \ \ \ \ \ \ \ \ \ \ \ \ \ \ \ \ \ \ \ \ \ \ \ \ \ \ \ \nonumber\\-\frac{\bar{g}_{jm}}{\tilde{H}(y)}\partial_{k}\!\left[e^{2\sigma}\tilde{H}(y)\bar{g}^{kl}\bar{g}^{ji}\left(\partial_{[l}\mathcal{A}_{i]}^{\ \ \mu_{1}..\mu_{p-1}}+\partial^{[\mu_{1}}\mathcal{A}^{\mu_{2}..\mu_{p-1}]}_{\ \ \ \ \ \ \ \ \ \ li}\right)\right]=0,\label{EQ-3-3-07}\\
\partial_{\rho}\!\left[\sqrt{-\hat{g}}\left(\partial^{[j}\mathcal{A}^{k]\rho\mu_{1}..\mu_{p-2}}+\partial^{[\rho}\mathcal{A}^{\mu_{1}..\mu_{p-2}]jk}\right)\right]=0.\label{EQ-3-3-08}
\end{eqnarray}
In the above equation we have defined $\hat{g}$ and $\bar{g}$ as the determinants of $\hat{g}_{\mu\nu}(x)$ and $\bar{g}_{jk}(y)$ respectively. We also defined $\tilde{H}(y)\equiv e^{[d-2p]\sigma(y)}\sqrt{\bar{g}(y)}$ and the index contractions in Eqs. (\ref{EQ-3-3-06})-(\ref{EQ-3-3-08}) are performed by using $\hat{g}_{\mu\nu}(x)$ or $\bar{g}_{jk}(y)$. 

In the above equations the effective fields are coupled. The simplification is a little bit more intricate than in the codimension one case and must be done carefully. First of all we will consider the gauge $\partial_{\mu_{1}}\left[\sqrt{-\hat{g}}\mathcal{A}^{\mu_{1}..\mu_{p}}\right]=0$ and $\mathcal{A}_{z\mu_{1}..\mu_{p-1}}(x,y)=0$. With this and by choosing $m=z$ in equation (\ref{EQ-3-3-07}) we get 
\begin{equation}\label{(p-1)form}
\partial_{w}\!\left[P(y)\left(-\partial_{z}\mathcal{A}_{w}^{\ \ \mu_{1}..\mu_{p-1}}+\partial^{[\mu_{1}}\mathcal{A}^{\mu_{2}..\mu_{p-1}]}_{\ \ \ \ \ \ \ \ \ \ wz}\right)\right]=0. 
\end{equation}
In the above equation $P(y)=e^{2\sigma}\tilde{H}(y)\bar{g}^{ww}\bar{g}^{zz}$ and we have used the fact that for all backgrounds considered here, the metric $\bar{g}_{jk}(y)$ is diagonal, with $\bar{g}_{zw}=0$. Now we perform the separation of variables
\begin{eqnarray}
\mathcal{A}_{\nu_{1}..\nu_{p}}(x,y)&=&A_{\nu_{1}..\nu_{p}}(x)\xi_{p}(y), \nonumber \\
\mathcal{A}_{\mu_{1}..\mu_{p-2}zw}(x,y)&=&A_{\mu_{1}..\mu_{p-2}}(x)\psi_{p}(y),\label{EQ-3-3-12} \\ 
\mathcal{A}_{\mu_{1}..\mu_{p-1}w}(x,y)&=&A_{\mu_{1}..\mu_{p-1}}(x)w_{p}(y).\nonumber
\end{eqnarray}
The index $p$ in the functions $\xi_{p}(y)$, $\psi_{p}(y)$ and $w_{p}(y)$ stress the solutions for a specific bulk $p$-form $\mathcal{A}_{N_{1}..N_{p}}$. With this, Eq. (\ref{(p-1)form}) gives us that
\begin{equation}\label{(p-1)form final}
A^{\mu_{1}..\mu_{p-1}}=f(y)\partial^{[\mu_{1}}A^{\mu_{2}..\mu_{p-1}]}, 
\end{equation}
where
$$
f(y)=\frac{\partial_{w}\!\left[P(y)\psi_{p}(y)\right]}{\partial_{w}\!\left[P(y)\partial_{z}w(y)\right]}.
$$
Eq. (\ref{(p-1)form final}) gives us the important results that
$$
\partial^{[\mu} A^{\mu_{1}..\mu_{p-1}]}=0.
$$
With this we get that the last term of Eq. (\ref{EQ-3-3-06}), the second of Eq. (\ref{EQ-3-3-07}) and the first of Eq. (\ref{EQ-3-3-08}) are all null.  The above equation also tell us that the $(p-1)$-form is a pure gauge and can be absorbed in the definition of our $(p-2)$-form. Therefore, finally we obtain that the equations (\ref{EQ-3-3-06})-(\ref{EQ-3-3-08}) are simplified to
\begin{eqnarray}
\frac{1}{\sqrt{-\hat{g}}}\partial_{\rho}\!\left[\sqrt{-\hat{g}}\mathcal{F}^{\rho\mu_{1}..\mu_{p}}\right]+\frac{e^{2\sigma}}{\tilde{H}(y)}\partial_{k}\!\left[\tilde{H}(y)\bar{g}^{kj}\partial_{j}\mathcal{A}^{\mu_{1}..\mu_{p}}\right]=0,\label{EQ-3-3-09}\\
\partial_{k}\!\left[e^{2\sigma}\tilde{H}(y)\bar{g}^{kl}\bar{g}^{ji}\partial_{[\mu_{1}}\mathcal{A}_{\mu_{2}..\mu_{p-1}]li}\right]=0,\label{EQ-3-3-10}\\
\partial_{\rho}\!\left[\sqrt{-\hat{g}}\partial^{[\rho}\mathcal{A}^{\mu_{1}..\mu_{p-2}]jk}\right]=0.\label{EQ-3-3-11}
\end{eqnarray}
Therefore, we reduced the system of equations (\ref{EQ-3-3-06})-(\ref{EQ-3-3-08}) to the system (\ref{EQ-3-3-09})-(\ref{EQ-3-3-11}) with only  two effective fields, namely, a $p$-form $\mathcal{A}_{\nu_{1}..\nu_{p}}$, and a $(p-2)$-form $\mathcal{A}_{\mu_{1}..\mu_{p-2}jk}$. Now, by using Eq. (\ref{EQ-3-3-12}) we obtain the separated equations
\begin{eqnarray}
\frac{1}{\sqrt{-\hat{g}}}\partial_{\rho}\!\left[\sqrt{-\hat{g}}F^{\rho\mu_{1}..\mu_{p}}(x)\right]&=&m^{2}A^{\mu_{1}..\mu_{p}}(x),\label{EQ-3-3-13}\\
\partial_{\rho}\!\left[\sqrt{-\hat{g}}\partial^{[\rho}A^{\mu_{1}..\mu_{p-2}]}(x)\right]&=&0,\label{EQ-3-3-14}\\
-\frac{e^{2\sigma}}{\tilde{H}(y)}\partial_{k}\!\left[\tilde{H}(y)\bar{g}^{kj}\partial_{j}\xi_{p}(y)\right]&=&m^{2}\xi_{p}(y),\label{EQ-3-3-15}\\
\partial_{k}\!\left[e^{2\sigma}\tilde{H}(y)\bar{g}^{kl}\bar{g}^{ji}\epsilon_{li}\psi_{p}(y)\right]&=&0,\label{EQ-3-3-16}
\end{eqnarray}
where we have used $\epsilon_{li}=-\epsilon_{il}$ and $\epsilon_{12}=1$. The above expressions are the effective equations for a $p$ and a $(p-2)$ forms, (\ref{EQ-3-3-13}) and (\ref{EQ-3-3-14}), with modes driven by equations  (\ref{EQ-3-3-15}) and (\ref{EQ-3-3-16}), respectively. We should point the curious fact that from the above equation we see that the effective $(p-2)$-form is always massless. 

Now, by substituting Eq. (\ref{EQ-3-3-12}) in the action (\ref{EQ-3-3-04}), we can see  that we get an effective Lagrangian, which is given by
\begin{eqnarray}\label{EQ-3-3-17}
\mathcal{L}_{(eff)}(x)=-\int d^{2}y\tilde{H}(y)\!\left[\xi_{p}^{2}e^{-2\sigma}\left(\frac{1}{2(p+1)!}F^{2}_{\mu_{1}..\mu_{p+1}}+\frac{1}{2p!}m^{2}A^{2}_{\mu_{1}..\mu_{p}}\right)\ \ \ \ \ \ \ \ \ \right.\nonumber\\+\left.\frac{1}{(p-1)!}\psi_{p}^{2}e^{2\sigma}\bar{g}^{11}\bar{g}^{22}F^{2}_{\mu_{1}..\mu_{p-1}}\right].
\end{eqnarray}
In the above Lagrangian $F_{\mu_{1}..\mu_{p-1}}=\partial_{[\mu_{1}}A_{\mu_{2}..\mu_{p-1}]}(x)$. Therefore, from a free massless bulk $p$-form field in codimension two, we obtained an effective free $p$-form confined with $\xi_{p}$ and a free massless $(p-2)$-form field confined with $\psi_{p}$ on the brane. 

In this manuscript, we will study only the localization of the zero-modes. To do this, we consider $m^{2}=0$ in equation (\ref{EQ-3-3-17}) and the effective Lagrangian is given by
\begin{eqnarray}\label{EQ-3-3-18}
\mathcal{L}^{(eff)}_{0}(x)=-\frac{1}{2(p+1)!}F^{2}_{\mu_{1}..\mu_{p+1}}(x)K_{1}-\frac{1}{(p-1)!}F^{2}_{\mu_{1}..\mu_{p-1}}(x)K_{2},
\end{eqnarray}
where $K_1$ and $K_2$ are given by
\begin{eqnarray}
K_{1}&=&\int d^{2}y\tilde{H}(y)\xi_{p}^{2}e^{-2\sigma}=\int d^{2}y e^{[d-2p-2]\sigma}\sqrt{\bar{g}}\xi_{p}^{2},\label{EQ-3-3-19}\\
K_{2}&=&\int d^{2}y\tilde{H}(y)\psi_{p}^{2}e^{2\sigma}\bar{g}^{11}\bar{g}^{22}=\int d^{2}y e^{[d-2p+2]\sigma}\sqrt{\bar{g}}\psi_{p}^{2}\bar{g}^{11}\bar{g}^{22}.\label{EQ-3-3-20}
\end{eqnarray}
From the above equations, the fields are said to be localized on the brane when the integrals $K_{a}$ are finite.

As said before, frequently in the literature the $(p-2)$-form is set to zero and only the integral $K_{1}$ is used\cite{Alencar07, Oda01, Oda03, Silva, Costa2013, Silva2011}.  Therefore these authors study only the confinement $p$-form component of the effective fields. Beyond this, in most of these cases the authors consider only the  constant solution of the zero-mode equation (\ref{EQ-3-3-15}). However, as shown in Ref. \cite{Duff} for codimension one models, the non-constant zero-mode solution obtained from (\ref{EQ-3-3-15}) has an important role to attain the consistency with Hodge duality. For codimension two this solution will not play that role. As will become clear, the crucial point is the presence of the effective $(p-2)$-form. 

To discuss the consequences of statement (i), we need to know what effective fields are confined for each braneworld model. We let this to the next subsection. Before this, we can develop some general consequences of statements (ii) for our codimension two scenario. For this, we will use the field configuration discussed above. From the separations of variables (\ref{EQ-3-3-12}) we get
\begin{eqnarray}
\mathcal{F}_{\mu_{1}..\mu_{p+1}}(x,y)&=&\xi_{p}(y)F_{\mu_{1}..\mu_{p+1}}(x),\label{EQ-3-3-21}\\
\mathcal{F}_{k\mu_{1}..\mu_{p}}(x,y)&=&\partial_{k}\mathcal{A}_{\mu_{1}..\mu_{p}}
=\partial_{k}\xi_{p}(y)A_{\mu_{1}..\mu_{p}}(x),\label{EQ-3-3-22}\\
F_{\mu_{1}..\mu_{p-1}12}(x,y)&=&\psi_{p}(y)\partial_{[\mu_{1}}\hat{A}_{\mu_{2}..\mu_{p-1}]}=\psi_{p}(y)\hat{F}_{\mu_{1}..\mu_{p-1}}(x).\label{EQ-3-3-23}
\end{eqnarray}
For the dual field we also get similar expressions. We must be careful and define $\psi_{d-p}$ and $\xi_{d-p}$ in the separation of variables for the dual fields. For the case of codimension two, Eq. (\ref{EQ-3-2-02}) can be written in components to gives us the independent relations
\begin{eqnarray}
\left(\star \mathcal{F}\right)^{\mu_{1}...\mu_{d-p-1}lm}=-\frac{(-1)^{(p+1)(d-p+1)}}{(p+1)!\sqrt{-g}}\varepsilon^{\mu_{1}...\mu_{d-p-1}\nu_{1}...\nu_{p+1}lm}\mathcal{F}_{\nu_{1}...\nu_{p+1}},\label{EQ-3-3-24}\\
\left(\star \mathcal{F}\right)^{\mu_{1}...\mu_{d-p}l}=-\frac{(-1)^{(p+1)(d-p+1)}}{(p+1)!\sqrt{-g}}\varepsilon^{\mu_{1}...\mu_{d-p}\nu_{1}...\nu_{p}lm}\mathcal{F}_{m\nu_{1}...\nu_{p}}.\label{EQ-3-3-25}
\end{eqnarray}
It is easy to see that the relation between the brane Greek index satisfy the Hodge duality transformation prescription, and we can define the Levi-Civita on the brane as $\epsilon^{\mu_{1}..\mu_{d}}\equiv\varepsilon^{\mu_{1}..\mu_{d}12}$. Therefore, the Hodge duality on the bulk, provided by (\ref{EQ-3-2-02}), is preserved and provided on the brane by (\ref{EQ-3-3-24}). Now, by using Eqs. (\ref{EQ-3-3-21})-(\ref{EQ-3-3-23}), the above equations give us
\begin{eqnarray}
\frac{\psi_{d-p}(y)\bar{g}^{11}\bar{g}^{22}\sqrt{\bar{g}}}{e^{2(d-p-1)\sigma}}(\star F)^{\mu_{1}..\mu_{d-p-1}}_{\ \ \ \ \ \ \ \ \ \ \ \ 12}(x)&\propto &-\frac{\xi_{p}(y)}{e^{d\sigma}}\frac{1}{\sqrt{-\hat{g}}}\varepsilon^{\mu_{1}..\mu_{d-p-1}\nu_{1}..\nu_{p+1}12}F_{\nu_{1}...\nu_{p+1}}(x),\label{EQ-3-3-26}\\
e^{-2(d-p)\sigma}\partial^{l}\xi_{d-p}(y)B^{\mu_{1}..\mu_{d-p}}(x)&\propto &-\frac{1}{\sqrt{-g}}\varepsilon^{\mu_{1}...\mu_{d-p}\nu_{1}...\nu_{p}lm}\partial_{m}\xi_{p}(y)A_{\nu_{1}..\nu_{p}}(x).\label{EQ-3-3-27}
\end{eqnarray}
From these, by considering $\psi_{d-p}$ and $\xi_{p}$ non-zero, we can factor out the function of the extra dimensions in the expression (\ref{EQ-3-3-26}). In doing this, we finally obtain the main result provided by statement (ii), namely,
\begin{eqnarray}\label{EQ-3-3-28}
\psi_{d-p}(y)\bar{g}^{11}\bar{g}^{22}\sqrt{\bar{g}}e^{-2(d-p-1)\sigma}\propto\xi_{p}(y)e^{-d\sigma}.
\end{eqnarray}

\begin{figure}
\centering
\includegraphics[scale=0.8,trim = 4.3cm 13.3cm 4.3cm 13.3cm,clip]{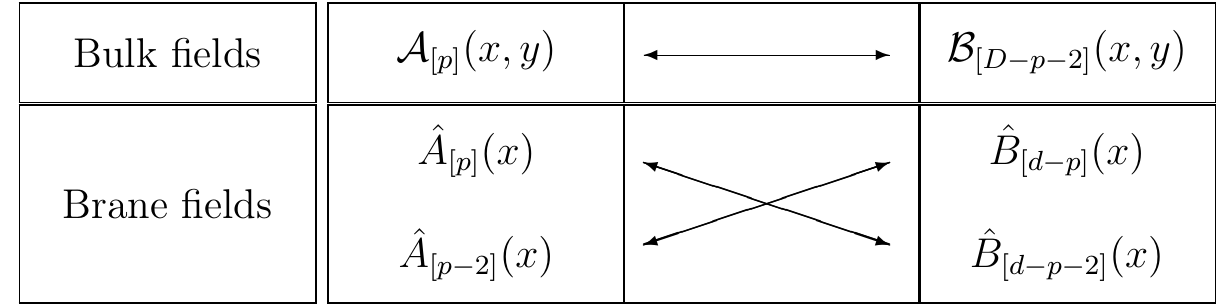}
\caption{Equivalency relation (indicated by the arrows) between the bulk fields, and also, between the effective brane fields after the dimensional reduction.}\label{Figura-01}
\end{figure}

The consequences of equation (\ref{EQ-3-3-28}) deserves some discussion. First, the function $\psi_{d-p}(y)$, obtained from the components $B_{\nu_{1}..\nu_{d-p-2}12}$, and $\xi_{p}(y)$ obtained for $A_{\nu_{1}..\nu_{p}}$ must satisfy the equations of motion (\ref{EQ-3-3-15}) and (\ref{EQ-3-3-16}). In general, the equations of motion provides us with two solutions for each value of $p$. However, due to the Hodge duality, we are not completely free to choose which one to use. Namely, if we choose one of the solutions $\xi_{p}(y)$ to confine the $p$-form, the only other consistent solution to its dual will be fixed by (\ref{EQ-3-3-28}). The figure (\ref{Figura-01}) shows a schematic picture of this equivalency relation between the bulk fields, and also between their components (effective fields on the brane). To discuss the above results in more details,  we will apply it to the models reviewed in section (\ref{Sec-2}). 

\subsection{Application}\label{Subsec-3-4}

To discuss the above results in a more practical setting, let us apply them to those braneworld models presented in section (\ref{Sec-2}). In doing this, we will also discuss the results found previously in the literature and its disagreement with the Hodge duality. We will use the notation [$p$-form, $(d-p)$-form] for the dual pair.

\begin{itemize}
\item[\hypertarget{(3A)}{(3A)}] Let us start by discussing the results of the previous subsection to the braneworld model presented in the case \hyperlink{(2A)}{(2A)} of section (\ref{Sec-2}). For this case, we reviewed the results obtained in Refs. \cite{Oda01, Oda03} for the scalar and the vector fields, and after generalized for a $p$-form in Ref. \cite{Alencar07}. As we showed, these references consider only the components $A_{\nu_{1}..\nu_{p}}$ nonzero and the zero-mode localization is attained with the constant solution of (\ref{EQ-2A-10}). For the local defect we have seen that the localization condition gives us 
\begin{equation}\label{localdefectcondition1}
p<\frac{(d-1)}{2}.
\end{equation}

For the global defect the condition is
\begin{equation}
-\frac{|\Lambda|}{\kappa^{2}_{D}}<t_{\theta}<\frac{(3-2p)|\Lambda|}{2(1+p)\kappa^{2}_{D}}.\label{EQ-3-4A-01}
\end{equation}
Thus, for the global defect, any $p$-form field can be confined by a suitable choice of the parameter $t_{\theta}$ \cite{Alencar07}. Now, let us make it clear where this approach fails with Hodge duality.

Let us apply the statement (i) and (ii) obtained in subsections (\ref{Subsec-3-2}) directly to the above confined $p$-form. Let us consider each case separately:
\begin{itemize}
\item {\bf The local defect}

Statement (i) tell us  that the confinement must include the bulk $p$-form and its bulk dual $(d-p)$-form. For the local defect, condition (\ref{localdefectcondition1}) clearly is in contradiction with this. Therefore, the solution is not consistent with Hodge duality.

\item {\bf The global defect} 

In this case, for $t_{\theta}\neq0$, all values of $p$ can be localized by using (\ref{EQ-3-4A-01}), therefore, the statement (i) can be satisfied. Now, from the statement (ii), the Hodge duality must be preserved on the brane. In other words, the bulk dual pair [$p$-form, $(d-p)$-form] must be found on the brane as the brane dual pair [$p$-form, $(d-p-2)$-form]. We discussed previously that this cannot be satisfied only with the components $A_{\mu_{1}...\mu_{p}}$. This can be verified by looking the equation (\ref{EQ-3-3-24}), which cannot be satisfied when the $A_{\mu_{1}...\mu_{p-2}lm}$ are zero. Since this is used in Ref. \cite{Alencar07}, the approach is inconsistent with the Hodge duality because it is not preserved after the dimensional reduction.  
\end{itemize}

To solve the above inconsistencies we must consider the $A_{\mu_{1}...\mu_{p-2}lm}$ component. Therefore we will use the approach presented in the previous subsection, which is a more complete description of this system. In doing this, let us start from equations of motion (\ref{EQ-3-3-15}) and (\ref{EQ-3-3-16}) for the massless modes. By using the metric (\ref{EQ-2A-02}), namely,
$$ds^{2}=e^{-2kr}\eta_{\mu\nu}dx^{\mu}dx^{\nu}+e^{-2(k-\delta)r}d\theta^{2}+dr^{2}$$
and by considering only the zero-mode $s$-state ($m,l=0$), Eqs. (\ref{EQ-3-3-15}) and (\ref{EQ-3-3-16}) can be written as
\begin{eqnarray}
\partial_{r}\!\left[e^{-(d-2p+1)kr+\delta r}\partial_{r}\xi_{p,0}(r)\right]&=&0,\label{EQ-3-4A-02}\\
\partial_{r}\!\left[e^{-(d-2p+1)kr-\delta r}\psi_{p}(r)\right]&=&0.\label{EQ-3-4A-03}
\end{eqnarray}
Here, unlike the discussion in Ref. \cite{Alencar07}, we have a new equation, namely, Eq. (\ref{EQ-3-4A-03}). As said above it will be crucial to attain the consistency with HD. From these equations, we get the solutions
\begin{eqnarray}
&\xi_{p,0,(1)}(r)=c_{1},\ \ \ \ \xi_{p,0,(2)}(r)=c_{2}e^{(d-2p+1)kr-\delta r},&\label{EQ-3-4A-04}\\
&\psi_{p}(r)=c_{3}e^{(d-2p+1)kr+\delta r}.&\label{EQ-3-4A-05}
\end{eqnarray}
Again, unlike the Ref. \cite{Alencar07}, we obtained the non-constant solution $\xi_{p,0,(2)}$ which the Ref. \cite{Alencar07} does not consider. Finally, we can put these solutions in the equations (\ref{EQ-3-3-19}) and (\ref{EQ-3-3-20}) to get
\begin{eqnarray}
K_{1}=R_{0}\int drd\theta e^{-[d-2p-1]kr+\delta r}\xi_{p,0}^{2}(r),\label{EQ-3-4A-06}\\
K_{2}=R_{0}\int drd\theta e^{-[d-2p+1]kr-\delta r}\psi_{p}^{2}(r).\label{EQ-3-4A-07}
\end{eqnarray}
From the above equations, the values of $p$ for which the $p$-form is localized can be obtained. For this we must impose that the integrals $K_{a}$ are finite. Let us discuss case by case below.

By using the zero-mode solutions (\ref{EQ-3-4A-04}) in equation (\ref{EQ-3-4A-06}) we will have two possibilities. First, the condition previously obtained for the constant solution $\xi_{p,0,(1)}$ given by Eq. (\ref{EQ-3-4A-01}) and obtained in Ref. \cite{Alencar07}.  However, by considering the non-constant solution $\xi_{p,0,(2)}$, we get a new condition, obtained for us, and given by
\begin{equation}
(d-2p+3)k<\delta \hspace{0.5cm}\stackrel{\scriptscriptstyle d=4}{\longmapsto}\hspace{0.5cm}-\frac{(7-2p)|\Lambda|}{2(1-p)\kappa^{2}_{D}}<t_{\theta}.
\end{equation}
For the local defect $\delta=0$ and we get 
\begin{equation}\label{newlocalcondition}
p>\frac{(d+3)}{2}.
\end{equation}
For the global defect, just as before, the above condition must be supplemented by the condition (\ref{EQ-2A-05}), for gravity localization. Together, these conditions give us 
\begin{equation}
-\frac{|\Lambda|}{\kappa^{2}_{D}}<-\frac{(7-2p)|\Lambda|}{2(1-p)\kappa^{2}_{D}}<t_{\theta}<\frac{3|\Lambda|}{2\kappa^{2}_{D}}, \ \ \mbox{for}\ \ p\geq3.\label{EQ-3-4A-08}
\end{equation}
Therefore, for $\xi_{p,0,(2)}$,  the range of $t_{\theta}$ in (\ref{EQ-3-4A-08}) is valid only for $p\geq3$. This is very different of when we use the constant solution, where we could localize any $p$-form.

Finally, we can analyze the localization of the effective $(p-2)$-form. By replacing (\ref{EQ-3-4A-05}) in the equation (\ref{EQ-3-4A-07}), it will be finite when $(d-2p+1)k<-\delta$. For the local defect this gives us 
\begin{equation}\label{(p-2)localcondition}
p>\frac{(d+1)}{2}.
\end{equation}
For the global defect we must use (\ref{EQ-2A-05}) to get, for $d=4$
\begin{equation}
-\frac{|\Lambda|}{\kappa^{2}_{D}}<t_{\theta}<-\frac{(5-2p)|\Lambda|}{2(5-p)\kappa^{2}_{D}}.\label{EQ-3-4A-09}
\end{equation}
Since 
$$
-\frac{|\Lambda|}{\kappa^{2}_{D}}<-\frac{(5-2p)|\Lambda|}{2(5-p)\kappa^{2}_{D}}\leq\frac{3|\Lambda|}{2\kappa^{2}_{D}}, \ \ \mbox{for}\ \ p\in\{0,4\}
$$
we always can find a value of $t_{\theta}$ to localize any $(p-2)$-form. This is very similar to the case of a constant solution of the $p$-form in the global defect.  

Thus, if we begin with a $p$-form $A_{N_{1}..N_{p}}$ in the bulk, we can have an effective $p$-form and a $(p-2)$-form  componentes, $A_{\mu_{1}..\mu_{p}}$ and $A_{\mu_{1}..\mu_{p-2}r\theta}$. In view of the above new results, let us discuss again the statement (i) and (ii).

{\bf Hodge duality}
 
As we showed in subsection (\ref{Subsec-3-3}), the Hodge duality was used to obtain a more complete description of the $p$-form field localization. In order to get this, we had to include the non-constant solution $\xi_{p,0,(2)}$ (Eq.  (\ref{EQ-3-4A-04})) for the $p$-form and also the $(p-2)$-form with solution $\psi_{p}$. Both solutions have not been found in Ref. \cite{Alencar07}. Now, we will show that the Hodge duality is present when these solutions are included. In doing this, we will consider $d=4$ and discuss the local and global cases separately:
\begin{itemize}
\item {\bf The local defect}

The effective $p$-form components $A_{\mu_{1}..\mu_{p}}$ will be localized if conditions (\ref{localdefectcondition1}) or (\ref{newlocalcondition}) are satisfied. Already for the effective $(p-2)$-form $A_{\mu_{1}..\mu_{p-2}r\theta}$, the confinement is attained by imposing the condition (\ref{(p-2)localcondition}). Condition (\ref{(p-2)localcondition}) is essential because it includes among the confined fields the components $A_{\mu_{1}..\mu_{p-2}r\theta}$ of a bulk $3$-form which is not allowed by (\ref{newlocalcondition}). This will be of crucial relevance to our present analysis. Therefore, for  $d=4$ the confined bulk $p$-form are those with $p=0,1$, localized with $\xi_{p,0,(1)}$, $p=4$ localized with $\xi_{p,0,(2)}$ or $\psi_{p}$ and $p=3$ localized with $\psi_{p}$. Remember that the bulk $p$-form confined with $\xi_{p,0}$ are found in the brane as $p$-form, but that confined with $\psi_{p}$ are found in the brane as $(p-2)$-form. With this, we can verify the statements (i) and (ii) obtained in subsection (\ref{Subsec-3-2}). From the statement (i), we get that the localization must include the bulk dual pair [$p$-form, $(4-p)$-form] simultaneously. By using the confined fields, we can relate them to get the bulk dual pairs [$0$-form, $4$-form] and [$1$-form, $3$-form]. These are all the confined fields for this model ($d=4$ and $t_{\theta}=0$), thus, the localization of these fields does not present any inconsistency with the statement (i). Here, we would like to stress that the bulk dual pair [$1$-form, $3$-form] presents a peculiarity. The bulk $3$-form field $A_{M_{1}M_{2}M_{3}}$ has the effective components $A_{\mu_{1}\mu_{2}\mu_{3}}$ and $A_{\mu r\theta}$, but only this last is confined, the components $A_{\mu_{1}\mu_{2}\mu_{3}}$ cannot be localized for this configuration ($d=4$ and $t_{\theta}=0$). Therefore, for the bulk $3$-form, the components $A_{\mu_{1}\mu_{2}\mu_{3}}$ must be considered zero. Now, let us verify the statement (ii). This statement says that the duality in the bulk must be preserved on the $3$-brane. Let us see this for the bulk dual pair [$0$-form, $4$-form]. As we showed above, the bulk $0$-form in confined with $\xi_{p=0,0,(1)}(r)$, thus, it is found on the brane as an effective $0$-form. However, the bulk $4$-form can be confined with $\xi_{p=4,0,(2)}(r)$ and $\psi_{p=4}(r)$. Thus, this field can be found on the brane as an effective $4$-form, by using the solution $\xi_{p=4,0,(2)}(r)$ and an effective $2$-form, by using the solution $\psi_{p=4}(r)$. The Hodge duality on the $3$-brane ($4$ dimensions) must relate the $0$-form and the $2$-form. Therefore, we must consider the localization of the bulk dual pair [$0$-form, $4$-form] with the solutions $\xi_{p=0,0,(1)}(r)$ and $\psi_{p=4}(r)$, which provide us with the effective dual pair [$0$-form, $2$-form]. About the effective $4$-form confined with $\xi_{p,0,(2)}(r)$, it is not a dynamical field in a $3$-brane. Now, let us verify the consistency of this discussion with equation (\ref{EQ-3-3-28}), which, for the above configuration ($d=4$ and $\delta=0$), gets
\begin{eqnarray}\label{EQ-3-4A-10}
\psi_{4-p}(r)\propto\xi_{p,0}(r)e^{-(3-2p)kr}.
\end{eqnarray}
As we said, for $p=0$, we must use $\xi_{p=0,0,(1)}(r)$ and equation (\ref{EQ-3-4A-10}) gets $\psi_{4}(r)\propto e^{-3kr}$. When we compare this with the solution (\ref{EQ-3-4A-05}) for $d=4$ e $\delta=0$, namely, $\psi_{p'}(r)=c_{3}e^{(5-2p')kr}$, we find that these solutions will be equal if $p'=4$. This result reinforce the statement (i), in the bulk the $0$-form is dual to the $4$-form. On the other hand, as the functions $\xi_{p=0,0,(1)}(r)$ and $\psi_{p=4}(r)$ are related to the pair of effective fields [$0$-form, $2$-form], the duality is preserved in the $3$-brane. Therefore, the statement (ii) is also satisfied. If we try to use the solution $\xi_{p,0,(2)}(r)$ in relation (\ref{EQ-3-4A-10}), we will get a function $\psi_{p}(r)$ that is not solution of (\ref{EQ-3-4A-03}). Therefore, the solution $\xi_{p,0,(2)}(r)$ cannot be used here. A similar conclusion is obtained for the bulk dual pair [$1$-form, $3$-form] which is found on the brane as the effective dual pair [$1$-form, $1$-form]. Therefore, these results correct the discussion performed in the reference  \cite{Alencar07} for the $p$-form fields and generalize the results found in Refs. \cite{Oda01, Oda03}, for the scalar and the vector fields, to include its dual fields.

\item {\bf The global defect} 

In this case the effective $p$-form components $A_{\mu_{1}..\mu_{p}}$ will be localized if conditions (\ref{EQ-3-4A-01}) or (\ref{EQ-3-4A-08}) are satisfied. Already for the effective $(p-2)$-form $A_{\mu_{1}..\mu_{p-2}r\theta}$, the confinement is attained by imposing the condition (\ref{EQ-3-4A-09}). From condition (\ref{EQ-3-4A-01}), as it was also showed in Ref. \cite{Alencar07}, any $p$ can be confined for this case. 
Therefore, the statement (i) is satisfied because it is always possible get a bulk dual pair [$p$-form, $(d-p)$-form]. 
Condition (\ref{EQ-3-4A-08}) tell us that only the $p$-form with $p\geq3$ are allowed with the non-constant solution $\xi_{p,0,(2)}$. An interesting point about this is that the $p$-form fields with $p\geq3$ can be confined with both solutions, $\xi_{p,0,(1)}$ or $\xi_{p,0,(2)}$. We are using `or' because the values of $t_{\theta}$ allowed for each solution are different and for a fixed value of $t_{\theta}$ the localization is carried out with $\xi_{p,0,(1)}$ or $\xi_{p,0,(2)}$ for those fields.  Therefore, by using only statement (i),  both solutions could be used to localize the effective fields with $p\geq3$. 

However, we have to verify the statement (ii). We can do this directly from equation (\ref{EQ-3-3-28}), namely, 
\begin{eqnarray}\label{EQ-3-4A-11}
\psi_{d-p}(r)\propto\xi_{p,0}(r)e^{-(d-2p-1)kr+\delta r}.
\end{eqnarray}
Here, we can discuss two possibilities. First, by using the constant solution $\xi_{p,0,(1)}(r)=c_{1}$, we get $\psi_{d-p}(r)\propto e^{-(d-2p-1)kr+\delta r}$. This solution is exactly that obtained in equation (\ref{EQ-3-4A-05}), namely, 
$$\psi_{p'}(r)=c_{3}e^{(d-2p'+1)kr+\delta r},$$
when we consider $p'=d-p$, which is the bulk dual of the $p$-form field. Thus, we have in the bulk the dual pair [$p$-form, $(d-p)$-form] which will be confined in the brane as the dual pair [$p$-form, $(d-p-2)$-form] preserving the duality on the brane. This result is valid for any $(d-1)$-brane. The other possibility is use the non-constant solution $\xi_{p,0,(2)}(r)$ of (\ref{EQ-3-4A-04}). In doing this, we get $\psi_{d-p}(r)\propto e^{2kr}$. This function is not a solution of (\ref{EQ-3-4A-03}) for any value of $p$. Thus, the non-constant solution $\xi_{p,0,(2)}(r)$ is not consistent with (\ref{EQ-3-4A-11}) and, therefore, must be eliminated. With this, we conclude the verification of the Hodge duality for our corrected results.   
\end{itemize}

An interesting point is that in both cases the HD eliminates the solution $\xi_{p,0,(2)}(r)$ which, until now, had no reason to be ruled out. As we saw above, this solution allows us to obtain some $p$-form fields confined, but it is not consistent with the Hodge duality. In this way, the duplicity due to the fact that the same $p$-form could be localized with $\xi_{p,0,(1)}(r)$ or $\xi_{p,0,(2)}(r)$, is solved by using this symmetry. These results also show that the approach used in Ref. \cite{Alencar07} for $t_{\theta}\neq 0$ is not consistent with Hodge duality. This because the statement (i) can be satisfied for the confined fields found in \cite{Alencar07}, but the statement (ii) cannot be satisfied for these fields. Finally it is important to stress the difference between  the codimension one and two cases. In codimension one, the non-constant solution is not excluded and in fact it is necessary to guarantee the HD. It is curious that here, on the contrary, this solution must be excluded by the same symmetry. In fact this is one of the most important results of this section. It will have important consequences when used together with the consistency with Einstein equation in the next section.


\item[\hypertarget{(3B)}{(3B)}] Now, let us discuss the braneworld models presented in the case \hyperlink{(2B)}{(2B)} of section (\ref{Sec-2}). In that review, Ref. \cite{Costa02} discuss the localization only of the vector field. Here, we will generalize this study for an arbitrary $p$-form field and, then, to discuss the Hodge duality. We should point that this has never been considered in the literature and therefore the below results are new. 

The action for the free $p$-form $\mathcal{A}_{N_{1}...N_{p}}$ is given by
\begin{eqnarray}\label{EQ-3-4B-01}
S^{(\mbox{mat})}=-\frac{1}{2(p+1)!}\int d^{d}xdrd\theta\sqrt{-g} \mathcal{F}_{N_{1}...N_{p+1}}\mathcal{F}^{N_{1}...N_{p+1}},
\end{eqnarray}
where $\mathcal{F}_{N_{1}...N_{p+1}}=\partial_{[N_{1}}\mathcal{A}_{N_{2}...N_{p+1}]}$ and $d$ is the brane dimension. Just like the previous case, the next mathematical steps are similar to those presented in subsection (\ref{Subsec-3-3}). Therefore, we can to start from the equations of motion (\ref{EQ-3-3-15}) and (\ref{EQ-3-3-16}). By using the metric (\ref{EQ-2B-01}), namely,
$$ds^{2}=e^{-kr+\tanh(kr)}\left[dx^{\mu}dx_{\mu}+\beta^{2}(a,r)d\theta^{2}\right]+dr^{2},$$
the equations (\ref{EQ-3-3-15}) and (\ref{EQ-3-3-16}) for the zero-mode (s-state) are given by
\begin{eqnarray}
\partial_{r}\left[e^{\frac{1}{2}(d-2p+1)[-kr+\tanh(kr)]}\tanh(kr)\partial_{r}\xi_{p,0}(r)\right]=0,\label{EQ-3-4B-02}\\
\partial_{r}\!\left[e^{\frac{1}{2}(d-2p+1)[-kr+\tanh(kr)]}\tanh^{-1}(kr)\psi_{p}(r)\right]=0.\label{EQ-3-4B-03}
\end{eqnarray}
Where $\xi_{p,0}$ is related to the components $\mathcal{A}_{\mu_{1}...\mu_{p}}$ (effective $p$-form) and $\psi_{p}$ is related to the components $\mathcal{A}_{\mu_{1}...\mu_{p-2}r\theta}$ (effective $(p-2)$-form). The solutions to the above equations are given by
\begin{eqnarray}
\xi_{p,0,(1)}(r)&=&c_{1},\label{EQ-3-4B-06}\\
\xi_{p,0,(2)}(r)&=&c_{2}\int^{r} \frac{e^{-\frac{\left(d-2p+1\right)}{2}\left[-kr'+\tanh(kr')\right]}}{\tanh(kr')}dr',\label{EQ-3-4B-07}\\
\psi_{p}(r)&=&c_{3}e^{-\frac{1}{2}[d-2p+1](-kr+\tanh(kr))}\tanh(kr).\label{EQ-3-4B-08}
\end{eqnarray}

With the zero mode solutions we can obtain the integrals $K_{a}$ of the extra dimensions in Eqs. (\ref{EQ-3-3-19}) and (\ref{EQ-3-3-20}), which can be written as
\begin{eqnarray}
K_{1}\propto \int dr e^{\frac{1}{2}(d-2p-1)[-kr+\tanh(kr)]}\tanh(kr)\xi^{2}_{p,0}(r),\label{EQ-3-4B-04}\\
K_{2}\propto \int dr e^{\frac{1}{2}(d-2p+1)[-kr+\tanh(kr)]}\tanh^{-1}(kr)\psi_{p}^{2}(r).\label{EQ-3-4B-05}
\end{eqnarray}
Thus, the localized values of $p$ can be obtained. Lets us analyze case by case.

First let us consider the localization of the effective $p$-form, thus, for which values of $p$ the integral (\ref{EQ-3-4B-04}) is finite. For this, just as for the vector case considered in section \hyperlink{(2B)}{(2B)}, we need of the asymptotic behavior of the integrand of $K_1$. For the constant solution (\ref{EQ-3-4B-06}) we get, as before, that the integrand is regular for all $r$. Therefore the localization condition will come from the behavior at large $r$, which is given by 
$$
e^{\frac{1}{2}(d-2p-1)[-kr+\tanh(kr)]}\tanh(kr)\xi^{2}_{p,0,(1)}(r)\to e^{-\frac{1}{2}(d-2p-1)kr}.
$$
With this we see that $K_1$ is finite only for 
\begin{equation}\label{EQ-3-4B-09}
p<\frac{(d-1)}{2}.
\end{equation}
Therefore, for $d=4$ we get that this solution localizes the scalar and the vector fields.

The other possibility for the integral $K_{1}$ is to use the non-constant solution $\xi_{p,0,(2)}(r)$. Just as before, this solution is singular at the origin since we have $\xi_{p,0,(2)}(r\to 0)\propto \ln(kr)$. However the integrand of  $K_{1}$ is regular for $r\to 0$ since 
$$
e^{\frac{1}{2}(d-2p-1)[-kr+\tanh(kr)]}\tanh(kr)\xi^{2}_{p,0,(2)}(r)\to kr[\ln(kr)]^{2}.
$$
Thus, the convergence of the complete integral (\ref{EQ-3-4B-04}) with $\xi_{p,0,(2)}(r)$ is determined by its behavior in the limit $r\to \infty$. In this limit, the asymptotic solution for $\xi_{p,0,(2)}(r)$ is given by
$$\xi_{p,0,(2)}(r\to\infty)\propto e^{\frac{(d-2p+1)}{2}kr}.$$
Now, we use this in the equation (\ref{EQ-3-4B-04}) to get 
$$
e^{\frac{1}{2}(d-2p-1)[-kr+\tanh(kr)]}\tanh(kr)\xi^{2}_{p,0,(2)}(r)\to e^{\frac{1}{2}(d-2p+3)kr}.
$$
From this, we can find that the confined $p$-form fields are only that where
\begin{equation}\label{EQ-3-4B-10}
2p>d+3.
\end{equation}
With this we see that now the non-constant solution can be used to localize some fields. This is very  different of the case for the vector field considered in section \hyperlink{(2B)}{(2B)}. We should also point that, as we saw in section  (\ref{Subsec-3-1}), this non-constant solution was used by \cite{Duff} to guarantee the HD in the codimension one braneworld.

Finally, we can discuss the integral (\ref{EQ-3-4B-05}). By using the solution (\ref{EQ-3-4B-08}) in $K_{2}$, we get that the integrand is regular for all $r$. Therefore the localization condition is obtained from the behavior at large $r$, which is given by
$$
e^{\frac{1}{2}(d-2p+1)[-kr+\tanh(kr)]}\tanh^{-1}(kr)\psi_{p}^{2}(r)\to e^{\frac{1}{2}(d-2p+1)kr}.
$$
From the above equation we see that the fields are localized only for  
\begin{equation}\label{EQ-3-4B-11}
2p>d+1.
\end{equation}
With this, we get all the confined $p$-form fields for this braneworld model. Here, we can stress the allowed fields for the particular case of a $3$-brane ($d=4$). For this, we get the bulk $0$-form and the bulk $1$-form fields confined with $\xi_{p,0,(1)}(r)$, the bulk $4$-form confined with $\xi_{p,0,(2)}(r)$ or $\psi_{p}(r)$ and also a $3$-form confined with $\psi_{p}(r)$. The case discussed in Ref. \cite{Costa02} for the vector field, of course, is included in our results. Next, let us discuss the consistency of these results with the Hodge duality. \\

{\bf Hodge duality}

Ref. \cite{Costa02} studied the localization only of the vector field, thus, it is not possible to verify the Hodge duality for that study. First of all, since the vector field was localized, its dual must also be. For this discussion, we will consider only the case of a $3$-brane ($d=4$). Once again, keep in mind that the bulk $p$-form confined with $\xi_{p,0}(r)$ are found in the brane as a $p$-form, but the bulk $p$-form confined with $\psi_{p}(r)$ are found in the brane as $(p-2)$-form. 

Now, let us verify the statements (i) and (ii) of subsection (\ref{Subsec-3-2}):  

\begin{itemize} 
 
\item {\bf Statement (i)}

From the statement (i), we must have confined the bulk dual pair [$p$-form, $(4-p)$-form]. We saw in the last subsection that the fields with $p=0,1,3$ and $4$ are all localized. Therefore the statement (i) can be satisfied because, with these fields, we can build the bulk dual pairs [$0$-form, $4$-form] and [$1$-form, $3$-form]. 

\item {\bf Statement (ii)}

Now, let us verify the statement (ii), which says that the Hodge duality must be preserved on the brane. This discussion is similar to that performed in the case \hyperlink{(3A)}{(3A)}. Let us see this for the bulk dual pair [$1$-form, $3$-form]. As the bulk $1$-form is confined with $\xi_{p=1,0,(1)}$ and the bulk $3$-form is confined with $\psi_{p=3}(r)$, these fields will be found on the brane as two $1$-forms. In this way, the bulk dual pair [$1$-form, $3$-form] is found on the brane as the dual [$1$-form, $1$-form], preserving the Hodge duality after the localization. Thus, the statement (ii) is also satisfied. The solution $\xi_{p=3,0,(2)}(r)$ does not allow us to confine the effective $3$-form, therefore, for this field, it must be made equal to zero. Now, we need to verify if $\xi_{p=1,0,(1)}(r)$ and $\psi_{p=3}(r)$ satisfy the equation (\ref{EQ-3-3-28}) for $d=4$, given by
\begin{eqnarray}\label{EQ-3-4B-12}
\psi_{4-p}(r)\propto\xi_{p,0}(r)e^{\frac{(3-2p)}{2}[-kr+\tanh(kr)]}\tanh(kr).
\end{eqnarray}
By using $\xi_{p=1,0,(1)}(r)=c_{1}$ in the above equation, we get 
$$
\psi_{3}(r)\propto e^{\frac{1}{2}[-kr+\tanh(kr)]}\tanh(kr).
$$
However, the solution (\ref{EQ-3-4B-08}) for $d=4$ must be 
$$
\psi_{p'}(r)=c_{3}e^{-\frac{5-2p'}{2}[-kr+\tanh(kr)]}\tanh(kr).
$$
When we compare the above solutions we find that these solutions will be equal if $p'=3$. This confirm the statement (i) and, as the function $\xi_{p=1,0,(1)}(r)$ and $\psi_{p=3}(r)$ are related to the pair of effective fields [$1$-form, $1$-form], the statement (ii) is also satisfied. A similar conclusion is obtained for the bulk dual pair [$0$-form, $4$-form] which is found on the brane as the effective dual pair [$0$-form, $2$-form]. 
\end{itemize} 

Regarding the study performed in Ref. \cite{Costa02} for the vector field, only with this analysis, there are no reasons to invalidate their results. In fact we have included more fields in order to preserve Hodge duality. We should also point a fact that will be important in the next section. Also for this model, just like the previous case, the non-constant solution $\xi_{p,0,(2)}$ is ruled out. This function is eliminated for the $3$-form because the localization cannot be attained and for the $4$-form because the components $\mathcal{A}_{\mu_{1}..\mu_{4}}$ are not dynamical fields in $4$ dimensions.


\item[\hypertarget{(3C)}{(3C)}] In this third example, we will discuss the case \hyperlink{(2C)}{(2C)} of section (\ref{Sec-2}). As we showed in that section, Ref.  \cite{Costa2013} discussed only the confinement of the vector field. Thus, in order to include the Hodge duality, we will discuss below the localization of the $p$-form field in the background (\ref{EQ-2C-01}). Again, these are new results that can not be found in the literature. However, some steps will be omitted because the discussion is very similar to that performed in the case \hyperlink{(3B)}{(3B)}. We can start from the action
\begin{eqnarray}\label{EQ-3-4C-01}
S^{(\mbox{mat})}=-\frac{1}{2(p+1)!}\int d^{d}xdrd\theta \sqrt{-g}\mathcal{F}_{N_{1}...N_{p+1}}\mathcal{F}^{N_{1}...N_{p+1}}.
\end{eqnarray}
With this and by using the metric (\ref{EQ-2C-01}), the equations of motion (\ref{EQ-3-3-15}) and (\ref{EQ-3-3-16}) for the zero-mode (s-state) give us
\begin{eqnarray}
\partial_{r}\left[e^{(d-2p+1)\sigma(r)}\beta(a,r)\partial_{r}\xi_{p,0}(r)\right]=0,\label{EQ-3-4C-02}\\
\partial_{r}\!\left[e^{(d-2p+1)\sigma(r)}\beta^{-1}(a,r)\psi_{p}(r)\right]=0.\label{EQ-3-4C-03}
\end{eqnarray}
In the above equations, $2\sigma(r)=-kr+\tanh(kr)$ and $\beta(a,r)$ is defined in (\ref{EQ-2C-02}). The solutions to the equations of motion are given by 
\begin{eqnarray}
\xi_{p,0,(1)}(r)&=&c_{1},\label{EQ-3-4C-06}\\
\xi_{p,0,(2)}(r)&=&c_{2}\int^{r} \frac{e^{-\left(d-2p+1\right)\sigma(r')}}{\beta(a,r')}dr',\label{EQ-3-4C-07}\\
\psi_{p}(r)&=&c_{3}e^{-(d-2p+1)\sigma(r)}\beta(a,r).\label{EQ-3-4C-08}
\end{eqnarray}

Finally, by using this  in Eqs. (\ref{EQ-3-3-19}) and (\ref{EQ-3-3-20}) we arrive at
\begin{eqnarray}
K_{1}\propto \int dr e^{(d-2p-1)\sigma(r)}\beta(a,r)\xi^{2}_{p,0}(r),\label{EQ-3-4C-04}\\
K_{2}\propto \int dr e^{(d-2p+1)\sigma(r)}\beta^{-1}(a,r)\psi_{p}^{2}(r).\label{EQ-3-4C-05}
\end{eqnarray}
 
We believe that the reader already understood the procedure that we are using to discuss the localization of a $p$-form. In fact, the next steps are the same to that carried out previously for the cases \hyperlink{(3A)}{(3A)} and \hyperlink{(3B)}{(3B)}. Considering this, we will be more brief in this discussion. Below we consider case by case separately.

First, let us discuss the localization of the effective $p$-form field. In this case we have two solutions. By using the constant solution (\ref{EQ-3-4C-06}) in the integral $K_{1}$, this integral will be finite only for
\begin{equation}\label{EQ-3-4C-09}
p<\frac{(d-1)}{2}.
\end{equation}
Therefore, with this solution, only the scalar and the vector fields are localized on a $3$-brane. On the other hand, by using the non-constant solution (\ref{EQ-3-4C-07}), the integral $K_{1}$ will be finite for
\begin{equation}\label{EQ-3-4C-10}
p>\frac{(d+3)}{2}.
\end{equation}
To see this, we must perform an asymptotic analysis of the integral (\ref{EQ-3-4C-07}) and this is similar to that discussed in the case \hyperlink{(3B)}{(3B)}. Therefore, for $d=4$ we have that this solution localizes only the forms with $p\geq 4$.

Now let us consider the effective $(p-2)$-form field. In this case we have just one solution (\ref{EQ-3-4C-08}). With this the integral $K_{2}$ given in Eq. (\ref{EQ-3-4C-05}), will be finite when
\begin{equation}\label{EQ-3-4C-11}
p>\frac{(d+1)}{2}.
\end{equation}
For $d=4$, for example, we get that this solution localizes only the forms with $p\geq 3$. 

For the particular case $p=1$ and $d=4$ we obtain the result of reference \cite{Costa2013}. However, we obtained the other fields can be localized. We should point that our results are independent of the resolution parameter $a$ and, beyond this, are the same obtained in the case \hyperlink{(3B)}{(3B)}. Therefore, the next conclusions about the consistency will be very similar to those obtained in that case. Below, let us discuss the Hodge duality for these results.\\

{\bf Hodge duality}

Let us consider the statements (i) and (ii) obtained from Hodge duality transformation. Once again, we will consider only a $3$-brane ($d=4$):  

\begin{itemize} 
 
\item {\bf Statement (i)}

From the statement (i), we must have confined the bulk dual pair [$p$-form, $(4-p)$-form]. We saw in the last subsection that, in this configuration: it is possible to confine a bulk $0$-form and $1$-form with $\xi_{p,0,(1)}$, a bulk $4$-form with $\xi_{p,0,(2)}$ or $\psi_{p}$ and also, a bulk $3$-form with $\psi_{p}$. Therefore the fields with $p=0,1,3$ and $4$ are all localized. From the statement (i) must be possible to confined the bulk dual pair [$p$-form, $(4-p)$-form] on the $3$-brane. By looking the allowed bulk $p$-form above, the statement (i) is satisfied if we get the bulk dual pairs [$0$-form, $4$-form] and [$1$-form, $3$-form].

\item {\bf Statement (ii)}

Now, let us verify the statement (ii), which says that the Hodge duality must be preserved on the brane. For the bulk dual pair [$0$-form, $4$-form], the statement (ii) can be satisfied only if we confine bulk $4$-form with $\psi_{p}$ because, for this case, the bulk $4$-form is found on the brane as an effective $2$-form. Therefore, the bulk dual pair [$0$-form, $4$-form] is found on the $3$-brane as the effective dual pair [$0$-form, $2$-form]. Note that the bulk $4$-form confined with $\xi_{p,0,(2)}$ in found on the brane as an effective $4$-form, thus, as the brane has $4$ dimensions, this is not a dynamical field. In the same way, the bulk dual pair [$1$-form, $3$-form] will be found on the brane as the effective dual pair [$1$-form, $1$-form] because the $3$-form is confined with $\psi_{p}$. To conclude, let us verify the equation (\ref{EQ-3-3-28}) for $d=4$. For this model it is given by
\begin{eqnarray}\label{EQ-3-4C-12}
\psi_{4-p}(r)\propto\xi_{p,0}(r)e^{\frac{(3-2p)}{2}[-kr+\tanh(kr)]}\beta(a,r).
\end{eqnarray}
Let us apply it to for the scalar field. For this case we have  which it is confined with $\xi_{p=0,0,(1)}(r)=c_{1}$ and therefore we get 
$$\psi_{4}(r)\propto e^{\frac{3}{2}[-kr+\tanh(kr)]}\beta(a,r).$$
However the solution (\ref{EQ-3-4C-08}), for $d=4$, gives us 
$$
\psi_{p'}(r)=c_{3}e^{-\frac{5-2p'}{2}[-kr+\tanh(kr)]}\beta(a,r).
$$
By comparing them, we find that these solutions will be equal only if $p'=4$. Once again, this confirm the statements (i) and (ii). A similar conclusion is obtained for the bulk dual pair [$1$-form, $3$-form].

\end{itemize} 

Thus, this study is a generalization of that  one performed in Ref. \cite{Costa2013}. For this more comprehensive analysis, beyond the vector field, other fields where localized in order to keep the consistency with the Hodge duality. Once again, just like the two previous cases, the non-constant $\xi_{p,0,(2)}$ is ruled out. In section \hyperlink{(3A)}{(3A)} this solution could localize the field but was not consistent with Hodge duality. For the present case, the solution $\xi_{p,0,(2)}$  is eliminated for the $3$-form because the localization cannot be attained. For the $4$-form the reason is that the components $\mathcal{A}_{\mu_{1}..\mu_{4}}$ are not dynamical fields in $4$ dimensions.\\


\item[\hypertarget{(3D)}{(3D)}] Until now, we discussed only string-like or conifold braneworld models. However, these are not the only ones that we can discuss the consistency with Hodge duality. The last model considered here will be that discussed in Ref. \cite{Kaloper2004} for codimension two and generalized in Ref. \cite{Arkani02} for arbitrary codimension. In the codimension two case, the extra dimensions are in the range $y^{1},y^{2}\in(-\infty,\infty)$ and the $3$-brane is generated by intersecting of delta-like $4$-branes. The metric for this configuration can be written as
\begin{eqnarray}\label{EQ-3-4D-01}
ds^{2}=\frac{1}{(1+k_{1}|y^{1}|+k_{2}|y^{2}|)^{2}}\left[dx^{\mu}dx_{\mu}+(dy^{1})^{2}+(dy^{2})^{2}\right].
\end{eqnarray}
Just like the previous cases, the spacetime has an asymptotically AdS$_{6}$ feature. The constants $k_{1}$ and $k_{2}$ are related to a negative cosmological constant on the bulk. 

About the $p$-form fields localization, there are some studies in the literature about this topic. Namely, Ref. \cite{Flachi} discussed the localization of the scalar and the vector fields, however, the analysis was performed for the interacting case. In Ref. \cite{Freitas} the vector field localization was also discussed and, again, the analysis was performed for the interacting case. In fact, the reference \cite{Freitas} discuss the free case, but only as a particular case. For the case of a $p$-form field, Ref. \cite{Chen} studied the localization of this field for a metric similar to that in (\ref{EQ-3-4D-01}), considering a more general and undefined warp factor. In this same reference the author discuss the Hodge duality. However, as they do not define or specify the metric, the confined $p$-form field are not obtained explicitly. Beyond this, the Hodge duality transformation is discussed in \cite{Chen} for other context, and they do not use it as a consistency test. Thus, for none of the above references we can apply directly the results developed here. In this way, we will discuss below the free $p$-form field localization in the background (\ref{EQ-3-4D-01}) and after to discuss the Hodge duality.

For this discussion, we will start from an action for the $p$-form field $\mathcal{A}_{N_{1}...N_{p}}$ given by
\begin{eqnarray}\label{EQ-3-4D-02}
S^{(\mbox{mat})}=-\frac{1}{2(p+1)!}\int d^{d}xd^{2}y\sqrt{-g} \mathcal{F}_{N_{1}...N_{p+1}}\mathcal{F}^{N_{1}...N_{p+1}},
\end{eqnarray}
where $\mathcal{F}_{N_{1}...N_{p+1}}=\partial_{[N_{1}}\mathcal{A}_{N_{2}...N_{p+1}]}$. From this, the next steps are similar to that developed in the previous cases. Thus, we will skip for the separated equations of motion. By using the metric (\ref{EQ-3-4D-01}), the equations of motion (\ref{EQ-3-3-15}) and (\ref{EQ-3-3-16}) for the zero-mode give us
\begin{eqnarray}
\partial_{j}\left[e^{[d-2p]\sigma\left(y\right)}\partial^{j}\xi_{p,0}(y)\right]=0,\label{EQ-3-4D-03}\\
\partial_{k}\!\left[e^{[d-2p]\sigma\left(y\right)}\delta^{kl}\delta^{ji}\epsilon_{li}\psi_{p}(y)\right]=0.\label{EQ-3-4D-04}
\end{eqnarray}
In the above expressions, $\sigma(y)=-\ln\left(1+k_{1}|y^{1}|+k_{2}|y^{2}|\right)$. The solutions of the above equations of motion are given by
\begin{eqnarray}
\xi_{p,0,(1)}(y)&=&c_{0},\ \ \ \ \ \xi_{p,0,(2)}(y)=c_{1}e^{-[d-2p+1]\sigma(y)},\label{EQ-3-4D-07}\\
\psi_{p}(y)&=&c_{3}e^{-[d-2p]\sigma(y)}.\label{EQ-3-4D-08}
\end{eqnarray}
With the above solutions, the localization integrals in Eqs. (\ref{EQ-3-3-19}) and (\ref{EQ-3-3-20}) can be written as
\begin{eqnarray}
K_{1}=\int dy^{1}dy^{2}e^{[d-2p]\sigma(y)}\xi_{p,0}^{2},\label{EQ-3-4D-05}\\
K_{2}=\int dy^{1}dy^{2}e^{[d-2p]\sigma(y)}\psi_{p}^{2}.\label{EQ-3-4D-06}
\end{eqnarray}
From this, we can discuss the allowed (confined) values of $p$ in this scenario.

Let us begin with the localization of the effective $p$-form field. In this case we have two solutions (\ref{EQ-3-4D-07}). If we consider the constant solution, the integral $K_{1}$ will be finite for
\begin{equation}\label{EQ-3-4D-09}
p<\frac{(d-2)}{2}.
\end{equation}
With the non-constant solution from (\ref{EQ-3-4D-07}), $K_{1}$ will be finite for
\begin{equation}\label{EQ-3-4D-10}
p>\frac{(d+4)}{2}.
\end{equation}
Therefore, for $d=4$ we get that the constant solution localizes only the scalar field. For the non-constant solution we get that the fields with $p>5$ are localized. 

Now let us turn to the effective $(p-2)$-form. We can use the solution (\ref{EQ-3-4D-08}) in Eq. (\ref{EQ-3-4D-06}) to obtain that $K_2$ is finite for
\begin{equation}\label{EQ-3-4D-11}
p>\frac{(d+2)}{2}.
\end{equation}
Therefore, for $d=4$ we get that only the cases with $p>3$ are localized. 

Let us summarize the results, by considering the particular case of a $3$-brane ($d=4$). With the constant solution $\xi_{p,0,(1)}$ we can confine only a bulk $0$-form. With the non-constant solution $\xi_{p,0,(2)}$ we can confine only a bulk $p$-form where $p\geq5$. With the solution $\psi_{p}$ we can confine only a bulk $p$-form where $p\geq4$. As we already mentioned, previously, the $p$-form confined with $\xi_{p,0}$ are found on the brane as an effective $p$-form and those confined with $\psi_{p}$ are found on the brane as an effective $(p-2)$-form. Thus, by looking the above allowed values of $p$, we get that the bulk $p$-form where $p\geq5$ confined with $\xi_{p,0,(2)}$ will be identically zero on the brane. Therefore, we can rule out the non-constant solution. Therefore, the only dynamical fields which are localized are the $0$ and the $4$-form. Now, let us discuss the Hodge duality in this scenario.\\

{\bf Hodge duality}

Let us consider the statements (i) and (ii) obtained from Hodge duality transformation. Once again, we will consider only a $3$-brane ($d=4$):  

\begin{itemize} 
 
\item {\bf Statement (i)}

From the statement (i), we must have confined the bulk dual pair [$p$-form, $(4-p)$-form]. We saw in the last subsection that, in this configuration, it is possible to confine only the bulk $0$-form and $4$-form. 
By looking the allowed bulk $p$-form above, the statement (i) can be satisfied because we can build the bulk dual pair [$0$-form, $4$-form]. 

\item {\bf Statement (ii)}

From the statement (ii), the Hodge equivalence must be preserved on the brane. For the bulk dual pair [$0$-form, $4$-form], the statement (ii) is satisfied because bulk $4$-form is confined with $\psi_{p}$, then, this is found on the brane as an effective $2$-form. Therefore, the bulk dual pair [$0$-form, $4$-form] is found on the $3$-brane as the effective dual pair [$0$-form, $2$-form]. Thus, the above results are in accordance with Hodge duality, as expected. To conclude, let us verify the equation (\ref{EQ-3-3-28}) for $d=4$. This equation for the present case is given by
\begin{eqnarray}\label{EQ-3-4D-12}
\psi_{4-p}(y)\propto\xi_{p}(y)e^{(4-2p)\sigma(y)}.
\end{eqnarray}
From this, as the scalar field is confined with $\xi_{p=0,0,(1)}(y)=c_{0}$, we get 
$$\psi_{4}(y)\propto e^{4\sigma(y)}.$$
However the solution (\ref{EQ-3-4D-08}) for $d=4$ is given by
$$
\psi_{p'}(r)=c_{3}e^{-[4-2p']\sigma(y)}.
$$
By comparing the above equations, we find that the solutions will be equal only if $p'=4$. Once again, this confirm the statements (i) and (ii).

\end{itemize}

\end{itemize}

We finally end this section with some comments. We showed that, in order to agree with the Hodge duality, we must use the more comprehensive approach used in subsection (\ref{Subsec-3-3}) to confine the $p$-form fields. We applied it for various braneworld models. We also showed that the non-constant solution $\xi_{p,0,(2)}$ must be ruled out for all for all the codimension two scenarios considered here. This is very different of the result of Ref. \cite{Duff}, where this solution was necessary to keep the Hodge duality. For this dimensional configuration, the HD is guaranteed not by the non-constant solutions, but by the components $\mathcal{A}_{\mu_{1}..\mu_{p-2}lk}$. This difference will be of fundamental importance in the next section. Therefore, the results of this section are twofold: a) To increased the set of $p$-form fields that can be confined by using the Hodge duality, b) To cancel the non-constant solutions. Next this will be used when we discuss the consistency of localization with the Einstein's equations.

\section{Consistency Conditions: Einstein's Equations}\label{Sec-4}

In this section, we will study the Einstein's equations, which must be satisfied by the fields in order that the localization of its zero-modes are consistently performed. This kind of condition has been considered in Ref. \cite{Duff} and, recently, the present authors performed a similar study for fields of different spins \cite{Freitas02}. Here, we will apply this consistency condition to the case of $p$-form fields discussed in the previous sections.

\subsection{Einstein's equations}\label{Subsec-4-1} 

Frequently, we focused our attention in the action (\ref{EQ-3-3-04}), without worrying about the effects of the matter fields on the background metric. In fact, the procedure described in the last section is performed by considering the background bulk metric. Any backreaction effect of that matter field on the background was ignored. Inspired by the study performed in Ref. \cite{Duff}, let us now to study the effect of backreaction by considering the full action
\begin{equation}\label{EQ-4-1-01}
S=S_{\mbox{(grav)}}-\frac{1}{2(p+1)!}\int d^{d}xd^{2}y\sqrt{-g}\mathcal{F}_{M_{1}...M_{p+1}}\mathcal{F}^{M_{1}...M_{p+1}}.
\end{equation}
We will analyze under what conditions the background metric solution obtained only from $S_{\mbox{(grav)}}$ is consistent with the solutions obtained from equations of motion (\ref{EQ-3-3-05}). Here, $S_{\mbox{(grav)}}$ is the action which defines the background metric and it is given by the Eq. (\ref{EQ-1-02}). From Eq. (\ref{EQ-4-1-01}), we get the following equations of motion
\begin{equation}\label{EQ-4-1-02}
 G_{MN}+g_{MN}\Lambda=\kappa \left(T^{(b)}_{MN}+T^{(\mbox{mat})}_{MN}\right),
 \end{equation} 
where $T^{(b)}_{MN}$ is the energy-momentum tensor that generate the braneworld. The stress tensor $T_{MN}^{(\mbox{mat})}$ is related to the differential $p$-form field, and it is given by
\begin{eqnarray}\label{EQ-4-1-03}
T_{MN}^{(\mbox{mat})}=\frac{1}{p!}\mathcal{F}_{MM_{2}..M_{p+1}}\mathcal{F}_{N}^{\ \ M_{2}..M_{p+1}}-\frac{1}{2(p\!+\!1)!}g_{MN}\mathcal{F}_{M_{1}..M_{p+1}}\mathcal{F}^{M_{1}..M_{p+1}}.
\end{eqnarray}
Here, we are considering that the presence of the $p$-form does not change the {\it shape} of the bulk metric. This is important to ensure that gravity will keep localized. However, as this field can be localized, the metric at the brane must be modified from $\eta_{\mu\nu}$ to $\hat{g}_{\mu\nu}(x)$. This modification of the metric at the brane lead to the following changes for the quantities $G_{\mu\nu}$ and $G_{jk}$
\begin{equation}\label{EQ-4-1-04}
G^{(vacuum)}_{\mu\nu}(y)=-\eta_{\mu\nu}f(y)\to G_{\mu\nu}(x,y)=\hat{G}_{\mu\nu}(x)-\hat{g}_{\mu\nu}(x)f(y)
\end{equation}
and 
\begin{equation}\label{EQ-4-1-05}
G^{(vacuum)}_{jk}(y)=-\eta_{jk}\tilde{f}(y)\to G_{jk}(x,y)=-\frac{1}{2}e^{-2\sigma}\bar{g}_{jk}\hat{R}(x)-\bar{g}_{jk}\tilde{f}(y).
\end{equation}
The quantity $T^{(b)}_{MN}$ also changes as $$T^{(b)}_{\mu\nu}(y)=\eta_{\mu\nu}h(y)\to T^{(b)}_{\mu\nu}(x,y)=\hat{g}_{\mu\nu}(x)h(y)$$
and $T^{(b)}_{jk}(y)$ does not change. Thus, from the components $(\mu,\nu)$ of the EE (\ref{EQ-4-1-02}), we get\footnote{This calculus can be found in more details in Ref. \cite{Freitas02}.} 
\begin{eqnarray}\label{EQ-4-1-06}
 \hat{G}_{\mu\nu}(x)=\kappa T_{\mu\nu}^{(\mbox{mat})}(x,y).
\end{eqnarray}
The result above was obtained by considering that the background metric is still valid. In this way, by consistency reasons, the energy-momentum tensor must satisfy 
\begin{eqnarray}\label{EQ-4-1-07}
T_{\mu\nu}^{(\mbox{mat})}(x,y)=T_{\mu\nu}^{(\mbox{mat})}(x).
\end{eqnarray}
Therefore, the above condition is necessary in order that the localization procedure to be consistent with background metric solution. In other words, the background metric, obtained from the action $S_{\mbox{(grav)}}$, can be used to study localization of fields. However, this is consistent with Einstein's equations only if the condition (\ref{EQ-4-1-07}) is satisfied. Otherwise, backreaction effects could destroy the background metric and therefore the entire model. The result (\ref{EQ-4-1-07}) is not new, it was obtained in Ref. \cite{Duff} for codimension one and it is independent of the codimension considered, as shown in Ref. \cite{Freitas02}. Below, we will apply this condition to the cases \hyperlink{(3A)}{(3A)}-\hyperlink{(3D)}{(3D)} discussed in previous section.

\subsection{Application}\label{Subsec-4-2}

To discuss the above results in a more practical setting, let us apply them to the previous mentioned codimension two braneworld models. To do this, we will need of the energy-momentum tensor for a free $p$-form field, namely,
\begin{eqnarray}\label{EQ-4-2-01}
T_{MN}^{(\mbox{mat})}=\frac{1}{p!}\mathcal{F}_{MM_{2}..M_{p+1}}\mathcal{F}_{N}^{\ \ M_{2}..M_{p+1}}-\frac{1}{2(p\!+\!1)!}g_{MN}\mathcal{F}_{M_{1}..M_{p+1}}\mathcal{F}^{M_{1}..M_{p+1}}.
\end{eqnarray}
From this, let us discuss the results presented in the review section (\ref{Sec-2}) and also those obtained by us in section (\ref{Sec-3}).

\begin{itemize}
\item[\hypertarget{(4A)}{(4A)}] Let us discuss first the results obtained in Refs. \cite{Oda01, Oda03, Alencar07} and presented in the case \hyperlink{(2A)}{(2A)} of the review section (\ref{Sec-2}). As we saw, for the scalar and the vector fields, the Refs. \cite{Oda01, Oda03} obtained the localization of these fields using the constant solution for $\xi_{p,0,(1)}$. After, in the Ref. \cite{Alencar07} the $p$-form localization was also attained  with this constant solution. For these cases, the stress tensor (\ref{EQ-4-2-01}) can be written as
\begin{eqnarray}\label{EQ-4-2A-00}
T_{\mu\nu}^{(\mbox{mat})}=e^{2pkr}\xi_{p,0}^{2}(r)\left[\frac{1}{p!}\hat{F}_{\mu \mu_{2}..\mu_{p+1}}\hat{F}_{\nu}^{\ \ \mu_{2}..\mu_{p+1}}-\frac{1}{2(p\!+\!1)!}\hat{g}_{\mu\nu}\hat{F}_{\mu_{1}..\mu_{p+1}}\hat{F}^{\mu_{1}..\mu_{p+1}}\right]\!\!(x).
\end{eqnarray}
Remember that for the Refs. \cite{Oda01, Oda03, Alencar07}, only the components $\mathcal{A}_{\mu_{1}..\mu_{p}}$ are nonzero. They also consider only the constant solution $\xi_{p,0,(1)}=c_{1}$. By using this we conclude that the above tensor is consistent with (\ref{EQ-4-1-07}) only for $p=0$. Therefore, from the results obtained in Refs. \cite{Oda01, Oda03, Alencar07}, only the scalar field can be consistently confined for this model. The vector field is ruled out by using this consistency condition. From these results, we could erroneously conclude that Einstein's equations (EE) are incompatible with Hodge duality. Because, the bulk dual of the $0$-form cannot be made consistent with EE (\ref{EQ-4-2A-00}). However, let us show below that this wrong conclusion is a consequence of the approach used in Refs. \cite{Oda01, Oda03, Alencar07}.

As we showed in last section, for the case \hyperlink{(3A)}{(3A)}, the solutions used to confine the $p$-form fields are given by
\begin{eqnarray}
&\xi_{p,0,(1)}(r)=c_{1},\ \ \ \ \xi_{p,0,(2)}(r)=c_{2}e^{(d-2p+1)kr-\delta r},&\label{EQ-4-2A-01}\\
&\psi_{p}(r)=c_{3}e^{(d-2p+1)kr+\delta r}.&\label{EQ-4-2A-02}
\end{eqnarray}
This enlarge the possibilities of fields which are consistent with Einstein's equation.  With these new solutions, and  by using the metric (\ref{EQ-2A-02}), the general energy-momentum tensor (\ref{EQ-4-2-01}) can be written, for the zero-mode, as
\begin{eqnarray}\label{EQ-4-2A-03}
T_{\mu\nu}^{(\mbox{mat})}(x,r)=e^{2pkr}\xi_{p,0}^{2}(r)\left[\frac{1}{p!}\hat{F}_{\mu \mu_{2}..\mu_{p+1}}\hat{F}_{\nu}^{\ \ \mu_{2}..\mu_{p+1}}-\frac{1}{2(p\!+\!1)!}\hat{g}_{\mu\nu}\hat{F}_{\mu_{1}..\mu_{p+1}}\hat{F}^{\mu_{1}..\mu_{p+1}}\right]\!\!(x)\nonumber\\
+e^{2(p-1)kr-2\delta r}\psi_{p}^{2}(r)\left[\frac{1}{(p-2)!}\hat{F}_{\mu \mu_{2}..\mu_{p-1}}\hat{F}_{\nu}^{\ \ \mu_{2}..\mu_{p-1}}-\frac{1}{2(p\!-\!1)!}\hat{g}_{\mu\nu}\hat{F}_{\mu_{1}..\mu_{p-1}}\hat{F}^{\mu_{1}..\mu_{p-1}}\right]\!\!(x).
\end{eqnarray}
Here, we already note the presence of other term which is not present in expression (\ref{EQ-4-2A-00}), and this term will provide the compatibility with HD as we will show below. 

From this expression, the consistency condition (\ref{EQ-4-1-07}) will be satisfied only if this stress tensor does not depend on the coordinate $r$. In this discussion, we will use directly the zero-mode solutions $\xi_{p,0}(r)$ and $\psi_{p}(r)$. Since we have two possibilities for $\xi_{p,0}(r)$ we will analyze each case separately. At the end we will  compare the results with the Hodge duality

\begin{itemize}
\item {\bf  The constant solution $\xi_{p,0,(1)}(r)$ }

The constant solution was considered in Refs. \cite{Oda01, Oda03, Alencar07}. However the authors did not consider the effective $(p-2)$-form. Here we must consider both $\xi_{p,0,(1)}(r)$ and also $\psi_{p}(r)$ in the stress tensor (\ref{EQ-4-2A-03}). In doing this, we get 
\begin{eqnarray}\label{EQ-4-2A-04}
T_{\mu\nu}^{(\mbox{mat})}(x,r)=e^{2pkr}c^{2}_{1}T_{\mu\nu}^{[p]}(x)+e^{2(d-p)kr}c^{2}_{3}T_{\mu\nu}^{[p-2]}(x),
\end{eqnarray}
where $T_{\mu\nu}^{[p]}(x)$ and $T_{\mu\nu}^{[p-2]}(x)$ are the expressions into the brackets of equation (\ref{EQ-4-2A-03}). From this, the consistency with EE can be obtained if the condition (\ref{EQ-4-1-07}) is satisfied. For the first term of the above equation, this is true only when $p=0$. This implies that the first term is $r$-independent and $T_{\mu\nu}^{[p-2]}(x)=0$ identically. This is the particular case obtained in Refs. \cite{Oda01, Oda03, Alencar07}. However, now we have a new possibility because the consistency with EE can also be obtained when we impose that $p=d$ in the second term of Eq. (\ref{EQ-4-2A-04}). This implies that the second term is $r$-independent and $T_{\mu\nu}^{[p]}(x)=0$ identically. An interesting point is that this new result provide the total agreement between the EE and the HD, since the $0$-form and the $d$-form are bulk duals. Beyond this, as the $d$-form is confined with $\psi_{p}(r)$, this field is found on the brane as an effective $(d-2)$-form, preserving the duality on the brane. The above results are independent of the parameter $t_{\theta}$. Therefore, for the particular case $t_{\theta}=0$, the conclusion will be the same.

\item {\bf  The non-constant solution $\xi_{p,0,(2)}(r)$ }

The second case that we can discuss is put the non-constant solution $\xi_{p,0,(2)}(r)$ and $\psi_{p}(r)$ in equation (\ref{EQ-4-2A-03}). In doing this, we get
\begin{eqnarray}\label{EQ-4-2A-05}
T_{\mu\nu}^{(\mbox{mat})}(x,r)=e^{2(d-p+1)kr-2\delta r}c^{2}_{2}T_{\mu\nu}^{[p]}(x)+e^{2(d-p)kr}c^{2}_{3}T_{\mu\nu}^{[p-2]}(x).
\end{eqnarray}
Therefore, once again, to obtain that the above expression is  consistency condition (\ref{EQ-4-1-07}) we have two possibilities. The first is by choosing the second term as independent of the extra dimensions, which can be attained $p=d$. This implies in $T_{\mu\nu}^{[p]}(x)=0$. The other possibility is to choosing $c_3=0$ and the first term of the above equation as independent of the extra dimensions. This imply that $(d-p+1)k=\delta$. By using the definition $\delta=2\kappa^{2}_{D}t_{\theta}/kd$ and $k$ given in Eq. (\ref{EQ-2A-03}), we get
$$t_{\theta}=\frac{d+1-p}{p}\frac{|\Lambda|}{\kappa^{2}_{D}}.$$
This condition is valid for any $d$ and it is into the allowed range (\ref{EQ-3-4A-08}). Therefore we get that the non-constant solution $\xi_{p,0,(2)}(r)$ can confine any $p$-form with $p\geq3$. Thus, there is no inconsistency between this value of $t_{\theta}$ and those allowed by the localization condition (\ref{EQ-3-4A-08}). 

\item {\bf Einstein Equation + Hodge duality conditions}

Now let us analyze the above results in the light of the Hodge duality. They are in contradiction with the previous discussion performed in the last section. As we showed in equation (\ref{EQ-3-3-28}), the presence of the function $\psi_{p}$ is crucial to get the consistency with HD. Beyond this, from the discussion about the Hodge duality, the non-constant solution $\xi_{p,0,(2)}(r)$ is not consistent with this symmetry and it was ruled out by this reason. In this way, to keep the consistency with the Hodge duality, the non-constant solution $\xi_{p,0,(2)}(r)$ will be eliminated also here. Thus, from this discussion, we find that,  among the various $p$ values that can be confined in this model, only the bulk scalar and the bulk $d$-form are consistent with Einstein equation and Hodge duality. These are exactly the equivalent fields related by HD in codimension two and, therefore, in total concordance with this symmetry. Any other field, such as the free vector claimed in Refs. \cite{Oda01, Oda03}, or other $p$-form claimed in \cite{Alencar07}, must be ruled out.
\end{itemize}

To end this subsection we would like to stress that the Einstein equation alone was not capable of ruling out the localization of the fields with $p>3$. This was obtained only when Hodge duality was also considered. Therefore, to impose both conditions is more restrictive than considering then alone.


\item[\hypertarget{(4B)}{(4B)}] This second example is that discussed in the case \hyperlink{(2B)}{(2B)} of the section (\ref{Sec-2}). In that review, we showed the results obtained for the vector field in Ref. \cite{Costa02}. Just like the previous case, the confinement of this field (zero-mode s-state) is attained with the constant solution. Let us verify the consistency of this result with the Einstein's equations. The stress tensor for the zero-mode s-state of the vector field is 
\begin{eqnarray}\label{EQ-4-2B-00}
T_{\mu\nu}^{(\mbox{mat})}=e^{-kr+\tanh(kr)}\xi_{0}^{2}(r)\left[\hat{F}_{\mu\rho}\hat{F}_{\nu}^{\ \ \rho}-\frac{1}{4}\hat{g}_{\mu\nu}\hat{F}_{\rho\lambda}\hat{F}^{\rho\lambda}\right]\!\!(x).
\end{eqnarray}
The consistency with EE is obtained when the above stress tensor is independent of $r$. By using the constant solution $\xi_{0}(r)=c_{1}$, we see that this cannot be attained. Therefore, the vector field localization discussed in \cite{Costa02} is not consistent with Einstein's equations. Here, even by considering the Hodge duality, the localization of this field is  inconsistent with EE, as we will show below.

Now, we will to apply the consistency condition (\ref{EQ-4-1-07}) to the model discussed in the case \hyperlink{(3B)}{(3B)} of last section. In doing this, let us use the energy-momentum tensor for the $p$-form field showed in equation (\ref{EQ-4-2-01}). This expression, for the zero-mode (s-state) of a general $p$-form field, is given by 
\begin{eqnarray}\label{EQ-4-2B-01}
T_{\mu\nu}^{(\mbox{mat})}(x,r)=e^{-2p\sigma(r)}\xi^{2}_{p,0}(r)T_{\mu\nu}^{[p]}(x)+\frac{e^{-2(p-1)\sigma(r)}}{\tanh^{-2}(kr)}\psi^{2}_{p}(r)T_{\mu\nu}^{[p-2]}(x),
\end{eqnarray}
where $2\sigma(r)=-kr+\tanh(kr)$. Just as before, from this expression, the consistency condition (\ref{EQ-4-1-07}) will be satisfied only if this stress tensor does not depend on the coordinate $r$. We will use directly the zero-mode solutions $\xi_{p,0}(r)$ and $\psi_{p}(r)$, and again have two possibilities for $\xi_{p,0}(r)$ which will be  analyzed each case separately. At the end we will compare the results with the Hodge duality.

\begin{itemize}
\item {\bf  The constant solution $\xi_{p,0,(1)}(r)$ }

The constant solution was considered in Ref. \cite{Costa02}. However the authors did not consider the effective $(p-2)$-form. Here we must consider both. However, as  we discussed previously, the allowed $p$-form fields for the constant solutions are only those where $p\leq 1$ and therefore the components $\hat{F}_{\mu_{1}..\mu_{p-1}}$ do not exist. With this the stress tensor in (\ref{EQ-4-2B-01}) gives us
\begin{eqnarray}\label{EQ-4-2B-02}
T_{\mu\nu}^{(\mbox{mat})}(x,r)=e^{-2p\sigma(r)}c^{2}_{1}T_{\mu\nu}^{[p]}(x).
\end{eqnarray}
Thus, we conclude that the relation (\ref{EQ-4-1-07}) will be satisfied only for $p=0$. The free vector field localization is really inconsistent with EE. Therefore the result of Ref. \cite{Costa02} is not consistent with Einstein equation. However the localization of the scalar field is.

\item {\bf  The non-constant solution $\xi_{p,0,(2)}(r)$ }

For the second case that we put the non-constant solution $\xi_{p,0,(2)}(r)$ and $\psi_{p}(r)$ in equation (\ref{EQ-4-2B-01}). In doing this, we get the energy-momentum tensor gives us
\begin{eqnarray}\label{EQ-4-2B-03}
T_{\mu\nu}^{(\mbox{mat})}(x,r)=e^{-2p\sigma(r)}\xi^{2}_{p,0,(2)}(r)T_{\mu\nu}^{[p]}(x)+c^{2}_{3}e^{-2(4-p)\sigma(r)}T_{\mu\nu}^{[p-2]}(x).
\end{eqnarray}
As before, the consistency of the above equation with  (\ref{EQ-4-1-07}) gives us two possibilities. The first is to choose the second term independent of $r$, which is attained for $p=4$. This implies that $T_{\mu\nu}^{[p]}(x)=0$ identically. For the other possibility we can see that, due to the complexity of $\xi_{p,0,(2)}(r)$, the first term can never be independent of $r$ by fixing the value of $p$. Therefore, just like the case \hyperlink{(4A)}{(4A)}, we conclude that in this case only the bulk $4$-form is consistent with Einstein's equation. 

\item {\bf Einstein Equation + Hodge duality conditions}

As we saw, the $p$-form allowed by EE consistency analysis, are only the bulk $0$-form and the bulk $4$-form, which are confined as effective $0$-form and $2$-form. These are exactly the equivalent fields related by the Hodge duality, therefore, the consistency with EE does not destroy the HD. Once again, the non-constant solution $\xi_{p,0,(2)}(r)$ is not important, however, now, this solution was eliminated independently of the Hodge duality.

\end{itemize}


\item[\hypertarget{(4C)}{(4C)}] The other model considered here will be that presented early in the case \hyperlink{(2C)}{(2C)} of the section (\ref{Sec-2}). In that review, we also showed the results obtained for the vector field in Ref. \cite{Costa2013}. Just like the previous cases, the confinement of this field (zero-mode s-state) is attained with the constant. Let us verify the consistency of this result with the Einstein's equations. The stress tensor for the zero-mode s-state of the vector field is 
\begin{eqnarray}\label{EQ-4-2C-00}
T_{\mu\nu}^{(\mbox{mat})}=e^{-kr+\tanh(kr)}\xi_{0}^{2}(r)\left[\hat{F}_{\mu\rho}\hat{F}_{\nu}^{\ \ \rho}-\frac{1}{4}\hat{g}_{\mu\nu}\hat{F}_{\rho\lambda}\hat{F}^{\rho\lambda}\right]\!\!(x).
\end{eqnarray}
The consistency with EE is obtained when the above stress tensor is independent of $r$. By using the constant solution $\xi_{0}(r)=c_{1}$, we see that this cannot be attained. Therefore, the vector field localization discussed in \cite{Costa2013} is not consistent with Einstein's equations.

Now, let us discuss the case \hyperlink{(3C)}{(3C)} presented early in the subsection (\ref{Subsec-3-4}). Again, to do this, we will need the stress tensor for the $p$-form field and, for the zero-mode (s-state), we get
\begin{eqnarray}\label{EQ-4-2C-01}
T_{\mu\nu}^{(\mbox{mat})}(x,r)=e^{-2p\sigma(r)}\xi^{2}_{p,0}(r)T^{[p]}_{\mu\nu}(x)
+\frac{e^{-2(p-1)\sigma(r)}}{\beta^{2}(a,r)}\psi_{p}^{2}(r)T^{[p-2]}_{\mu\nu}(x).
\end{eqnarray}
In the above equation, we will use directly the solutions $\xi_{p,0}$ and $\psi_{p}$. As before, let us analyze case by case.
\begin{itemize}
\item {\bf  The constant solution $\xi_{p,0,(1)}(r)$ }

The constant solution was considered in Ref. \cite{Costa2013}. However the authors did not consider the effective $(p-2)$-form. Here we must consider both. By using the solutions (\ref{EQ-3-4C-06}) and (\ref{EQ-3-4C-08}) for $\xi_{p,0,(1)}$ and $\psi_{p}$, we get
\begin{eqnarray}\label{EQ-4-2C-02}
T_{\mu\nu}^{(\mbox{mat})}(x,r)=e^{-2p\sigma(r)}c^{2}_{1}T^{[p]}_{\mu\nu}(x)
+c^{2}_{3}e^{-2(4-p)\sigma(r)}T^{[p-2]}_{\mu\nu}(x).
\end{eqnarray}
Thus, again we have to possibilities to reach the consistency with (\ref{EQ-4-1-07}). The first is to choose $p=0$. With this, the first term in the above stress tensor is $r$-independent and the term $T^{[p-2]}_{\mu\nu}(x)$ is identically zero. The other is to choose $p=4$. Now, the second term in (\ref{EQ-4-2C-02}) is $r$-independent and the term $T^{[p]}_{\mu\nu}(x)$ is identically zero. Therefore the localization of the vector field, performed in Ref. \cite{Costa2013}, is not consistent with Einstein equation. However the localization of the $0$ and $4$-form are. 

\item {\bf  The non-constant solution $\xi_{p,0,(2)}(r)$ }

The other possibility is use the non-constant solution $\xi_{p,0,(2)}$, instead the solution $\xi_{p,0,(1)}$. 
For this we substitute  the non-constant solution $\xi_{p,0,(2)}(r)$ and $\psi_{p}(r)$ in equation (\ref{EQ-4-2C-01}). In doing this, we get the energy-momentum tensor gives us
\begin{eqnarray}\label{EQ-4-2C-03}
T_{\mu\nu}^{(\mbox{mat})}(x,r)=e^{-2p\sigma(r)}\xi^{2}_{p,0,(2)}(r)T_{\mu\nu}^{[p]}(x)+c^{2}_{3}e^{-2(4-p)\sigma(r)}T_{\mu\nu}^{[p-2]}(x).
\end{eqnarray}
As before, the consistency of the above equation with  (\ref{EQ-4-1-07}) gives us two possibilities. First, the second term can be independent of $r$ if we choose $p=4$, which implies that $T_{\mu\nu}^{[p]}(x)=0$. However, it is not possible to consider  $e^{-2p\sigma(r)}\xi^{2}_{p,0,(2)}$ a constant for all $r$ by choosing the value of $p$. Thus, we conclude that only the $4$-form can be confined consistently with Einstein's equation.

\item {\bf Einstein Equation + Hodge duality conditions}

As we saw, the $p$-form allowed by EE consistency analysis, are only the bulk $0$-form and the bulk $4$-form, which are confined as effective $0$-form and two-form. These are exactly the equivalent fields related by the Hodge duality, therefore, the consistency with EE does not destroy the HD. Once again, the non-constant solution $\xi_{p,0,(2)}(r)$ is ruled out, however, now, this solution was eliminated independently of the Hodge duality.

\end{itemize}

Therefore, the result of Ref. \cite{Costa2013} is not consistent with Einstein Equation. Once again, the $p$-form allowed by EE consistency analysis are only the bulk $0$-form and the bulk $4$-form, which are confined as effective $0$-form and two-form. Therefore, the localization of these fields are in total concordance with the Hodge duality and the Einstein's equations. Also, for this case, the non-constant solution $\xi_{p,0,(2)}(r)$ is ruled out  and this is attained independently of the HD.


\item[\hypertarget{(4D)}{(4D)}] Finally, we can also discuss the consistency with Einstein's equations of case \hyperlink{(3D)}{(3D)}. For this case, the energy-momentum tensor can be written for the zero-mode as 
\begin{eqnarray}\label{EQ-4-2D-01}
T_{\mu\nu}^{(\mbox{mat})}(x,r)=e^{-2p\sigma(y)}\xi_{p,0}^{2}(y)T^{[p]}_{\mu\nu}(x)+e^{-2p\sigma(y)}\psi_{p}^{2}(y)T^{[p-2]}_{\mu\nu}(x).
\end{eqnarray}
Where $T^{[q]}_{\mu\nu}(x)$ is the effective stress tensor of the confined effective massless $q$-form. Thus, by using the solutions $\xi_{p,0,(1)}$ and $\psi_{p}$, the above equation gives
\begin{eqnarray}\label{EQ-4-2D-02}
T_{\mu\nu}^{(\mbox{mat})}(x,r)=e^{-2p\sigma(y)}c_{0}^{2}T^{[p]}_{\mu\nu}(x)+e^{-2(4-p)\sigma(y)}c_{3}^{2}T^{[p-2]}_{\mu\nu}(x).
\end{eqnarray}
From this, we get that the consistency with EE can be attained for $p=0$ or $p=4$. These results, once again, agree with the Hodge duality analysis. As we already mentioned, early the non-constant solution $\xi_{p,0,(2)}$ for this model does not allow us to confine a dynamical field on the brane.

\end{itemize}

\section{Conclusions}\label{Sec-5}

Until recently, the study about fields localization was performed considering only the finite integral argument. This argument states that a field is confined when the integral of the extra dimensions in the action is finite. In this context, the authors of Ref. \cite{Duff} discussed the free $p$-form field localization in codimension one models. However, they explored other aspects of the theory, namely, the Hodge duality (HD) and the Einstein's equations (EE). With this, they showed that a consistent localization must satisfy, beyond the finite integral, other conditions provided by HD and EE. For example, by using the Hodge duality, they showed that the localization of a free massless $p$-form must imply, necessarily, the localization of its bulk dual $(d-p-1)$-form. Inspired by the above reference, the present authors, in Ref. \cite{Freitas02}, discussed the consistency between the Einstein's equations and the localization of the scalar, vector and half spin fields. As a main conclusion, we obtained that the localization of these fields, attained in the literature, are not consistent with this new condition and must be ruled out. Reference \cite{Duff} did not consider the codimension two case and reference \cite{Freitas02} did not consider the consistency of $p$-forms with Einstein's equations. Thus, in this manuscript we propose to fill this gap.  

The Hodge duality provides an on-shell equivalency between a free massless $p$-form and a free massless $(D-p-2)$-form. From this, we obtain two general statements to guide us when applying to field localization: a) [statement (i)] for an arbitrary codimension two ($D=d+2$) model, the confinement must be possible for both free massless fields, the bulk $p$-form and its bulk dual $(d-p)$-form and b) [statement (ii)] this equivalence must be preserved for the effective fields over the $(d-1)$-brane. Based on these statements, we found some general results that must be applied to the localization of any free massless $p$-form in codimension two. The main consequence of statement (i) is that if some differential form is localized, its dual also must be and therefore the Hodge duality enlarges the set of confined fields. For example, since Refs. \cite{Oda01, Oda03, Costa02, Costa2013, Silva2011} find that the scalar and vector fields are localized, we necessarily must have that its duals also must be localized. However, we have shown that for this to be true we need of the components $\mathcal{A}_{\mu_{1}...\mu_{p-1}r\theta}$, which was not considered in Ref. \cite{Alencar07}, for example. With this at hand, we obtain the main consequence of statement (ii): the solutions to the zero-mode of a $p$-form must be linked to the zero-mode of its dual by the relation
\begin{eqnarray}\label{conclusion}
\psi_{d-p}(y)\bar{g}^{11}\bar{g}^{22}\sqrt{\bar{g}}e^{-2(d-p-1)\sigma}\propto\xi_{p}(y)e^{-d\sigma}.
\end{eqnarray}
Therefore, diverse of what is generally found in the literature, beyond to satisfy the mass equation, the solutions must satisfy the above constraint. The above relation is very general and is valid for any codimension two braneworld.

Next, in subsection (\ref{Subsec-3-4}), we applied the above general results for some $6$D braneworld models reviewed in section (\ref{Sec-2}). First, we considered the string-like brane model of subsection \hyperlink{(2A)}{(2A)}, with metric given by (\ref{EQ-2A-02}). In this background, the authors of Refs. \cite{Oda01, Oda03} discussed the localization of the free scalar and the free vector fields. Later, the authors of Ref. \cite{Alencar07} generalized this study for a free $p$-form and found the general localization condition. For the local and global defect, they are given respectively by
\begin{equation} \label{stringlike}
p<\frac{(d-1)}{2},\ \ \ -\frac{|\Lambda|}{\kappa^{2}_{D}}<t_{\theta}<\frac{(3-2p)|\Lambda|}{2(1+p)\kappa^{2}_{D}}.
\end{equation}
An important point for us is that, to reach the above conclusions, the authors considered the particular solution where the effective $(p-2)$ component is null, namely, $\mathcal{A}_{\mu_{1}...\mu_{p-2}r\theta}=0$. They also consider only the constant solution to the zero-mode equation $\mathcal{A}_{\mu_{1}...\mu_{p}}(x,y)=A_{\mu_{1}...\mu_{p}}(x)$. Let us first see the consequences of HD for the local case. In subsection \hyperlink{(3A)}{(3A)}, we revisit this result in the light of the Hodge duality. We showed that the results of those authors are not consistent with this symmetry. For example, the only fields localized in the local defect are those with $p<(d-1)/2$. As said above, HD demands that also the field with $p>(d+1)/2$ must also be localized. Here we find the first difference with the codimension one case, considered by Duff et al. in Ref \cite{Duff}. There, the authors found that this can be cured by including the non-constant solution for extra dimension of $\mathcal{A}_{\mu_{1}...\mu_{p}}(x,y)$, which provides the localization for the dual field. However, this cannot be made for our codimension two case. When we considered the non-constant solution we found condition (\ref{newlocalcondition}), given by $p>(d+3)/2$. What is going on? The point is that now we have the effective $(p-2)$-form. The solution (\ref{(p-2)localcondition}) for this component provides us exactly that the $p$-forms localized are the ones with $p>(d+1)/2$. Beyond this, the solutions also satisfy (\ref{conclusion}). Therefore, as said above, the HD demands that we cannot fix this field to zero. With this, we show explicitly that, beyond the scalar and vector fields of Refs. \cite{Oda01, Oda03} and for the local case in Ref. \cite{Alencar07}, its duals are localized. Let us turn to the global case, with $\delta\neq 0$. References Refs. \cite{Oda01, Oda03} show that the scalar field can be localized and Ref. \cite{Alencar07} show that, by choosing $t_{\theta}$, any $p$-form can be localized with the constant solution to the zero-mode. This is given by the second condition in Eq. (\ref{stringlike}). We could be leaded to think that, since any effective $p$-form field can be localized by choosing a value of $t_{\theta}$, HD is trivially respected. However, we have shown that with this choice the field and its dual do not satisfy condition (\ref{conclusion}). Even when we consider the non-constant solution, this cannot be obtained. Again, this is only possible if we include the effective $(p-2)$-form. We find that all the effective $p$ and $(p-2)$-forms are localized. However, they always form pairs of duals, and therefore both effective fields are necessary to guarantee HD. A further consequence of this is that, in fact, the non-constant solution does not have any dual. Therefore, this solution, even being capable of localizing some fields, must be excluded. With this we get the HD can also be used to exclude some zero-mode solutions. Next, we consider the models \hyperlink{(3B)}{(3B)} and \hyperlink{(3C)}{(3C)}, discussed by references \cite{Costa02, Costa2013, Silva2011}. The authors in these references only considered the localization of vector fields with the constant solution to the zero-mode equation. We generalized their results to include $p$-forms and considered also the effective $(p-2)$-form. We also considered the non-constant solution. With this we found analogous results as in the local defect of case \hyperlink{(3A)}{(3A)}. Namely, that beyond the scalar and vector fields, we must have its duals localized. However, this is obtained only when we include the components $\mathcal{A}_{\mu_{1}...\mu_{p-2}r\theta}=0$. We show that the dual solutions are linked by equation (\ref{conclusion}) and that the non-constant solutions must be excluded. We finally discussed another type of model, generated by the intersecting of delta-like branes \cite{Flachi, Arkani02, Kaloper2004}. For this model, case \hyperlink{(3D)}{(3D)}, the finite integral condition applied for the free $p$-form gives us, for a $3$-brane, only a bulk scalar field and a bulk $4$-form. Fortunately, these are exactly the Hodge dual fields on the bulk, and they are confined on the brane as the dual pair [$0$-form, $2$-form], in total agreement with the Hodge duality (statements (i) and (ii)).

In section \ref{Sec-4}, we discussed the consistency of the localization with Einstein's equations. For all the models, except for the local defect of case \hyperlink{(3A)}{(3A)}, we obtained the same conclusion: only the bulk $0$-form and its bulk dual $4$-form can be confined consistently with Einstein's equations. For the global defect we found that, beyond the scalar and its dual, the forms with $p\geq3$, localized with the non-constant solution, are consistent with EE. However, this conclusion is in contradiction with the previous discussion performed for the Hodge duality. The fact is that, from the discussion about the Hodge duality, the non-constant solution $\xi_{p,0,(2)}(r)$ is not compatible with this symmetry and it must be ruled out by consistency reasons. Therefore, despite being consistent with EE, the non-constant solutions, which localizes fields $p\geq3$, must be ruled out for the global defect case.

To conclude, we would like to stress the results obtained for the vector field. In Ref. \cite{Freitas02}, the present authors already has shown that the free vector field localization cannot be consistent for a large class of models. Here, we confirm this result by using different $6$D braneworld models. This is an important result, considering the variety of works in the literature which states, by using only the finite integral argument, that a free $U(1)$ gauge field can be localized in $6$D naturally \cite{Oda01, Oda03, Alencar07, Costa02, Costa2013, Silva2011}. Therefore, the localization of vector field in codimension two is reopened by result of the present manuscript. By using the HD, we found yet that, despite that the vector field is not localized, the scalar and its dual must be localized in all the above models. An interesting consequence of our result is that imposing both consistency conditions, with EE plus HD, is stronger than imposing just one. This show the interesting fact that the HD, a symmetry of the fields, was used to exclude zero-mode solutions of the equations of motion. This opens a new kind of analyzes that must be done for the different kinds of fields considered in braneworld scenarios. For example, if there may be other symmetries that can be used as consistency conditions in the study about fields localization. We should point that here we have considered just free fields and an on-shell duality of $p$-forms. However, for a consistent quantum theory, we should consider theories in which the above symmetry is valid of-shell. This has been considered recently for example in Ref. \cite{Boulanger} (and references therein). We expect that this and the generalization to other fields and symmetries will be presented by us in the near future.

\vfill
\section*{Acknowledgement}
The authors would like to thank Alexandra Elbakyan and sci-hub, for removing all barriers in the way of science. We acknowledge the financial support provided by the Conselho Nacional de Desenvolvimento Científico e Tecnológico (CNPq) and Fundação Cearense de Apoio ao Desenvolvimento Científico e Tecnológico (FUNCAP) through PRONEM PNE-0112-00085.01.00/16.

\end{document}